\documentclass[]{aa}  

\usepackage{changepage}
\usepackage{float}
\usepackage{graphicx}
\usepackage{subfig}
\usepackage{txfonts}
\usepackage{graphicx,rotating}
\usepackage{subfig}
\usepackage{txfonts}
\usepackage[T1]{fontenc}
\usepackage{epsfig}
\usepackage{natbib}
\usepackage{color}
\usepackage{natbib}
\usepackage{amssymb}  
\usepackage{tabularx}
\usepackage{comment}

\DeclareGraphicsExtensions{.pdf,.png,.jpg}
\bibpunct{(}{)}{;}{a}{}{,} 
\begin{document}

   \title{Molecular gas in radio galaxies in dense megaparsec-scale environments at $z=0.4-2.6$}

   \author{G. Castignani
          \inst{1,2,3}\fnmsep\thanks{e-mail: gianluca.castignani@epfl.ch}
          \and
          F. Combes\inst{2,3}
          \and
          P. Salom\'e\inst{2}
          \and
          C. Benoist\inst{4}
          \and
          M. Chiaberge\inst{5,6} 
          \and
          J. Freundlich\inst{7}
          \and
          G. De Zotti\inst{8}
          }

   \institute{Laboratoire d'astrophysique, \'{E}cole Polytechnique F\'{e}d\'{e}rale de Lausanne (EPFL), Observatoire de Sauverny, 1290 Versoix, Switzerland
              \and   
             Sorbonne Universit\'{e}, Observatoire de Paris, Universit\'{e} PSL, CNRS, LERMA, F-75014, Paris, France
               \and
             Coll\`{e}ge de France, 11 Place Marcelin Berthelot, 75231 Paris, France
             \and
             Universit\'{e} C\^{o}te d'Azur, Observatoire de la C\^{o}te d'Azur,  CNRS, Laboratoire Lagrange, Blvd de l'Observatoire, CS 34229, 06304 Nice cedex 4, France
             \and
             Space Telescope Science Institute, 3700 San Martin Dr., Baltimore, MD 21210, USA
             \and
             Johns Hopkins University, 3400 N. Charles Street, Baltimore, MD 21218, USA
             \and
             Centre for Astrophysics and Planetary Science, Racah Institute of Physics, The Hebrew University, Jerusalem 91904, Israel
             \and
             INAF-Osservatorio Astronomico di Padova, Vicolo dell'Osservatorio 5, I-35122 Padova, Italy
              }
             
                \date{Received 10 August 2018; Accepted 24 December 2018}
 
  \abstract
   {Low luminosity radio galaxies (LLRGs) typically reside in dense megaparsec-scale environments and are often associated with brightest cluster galaxies (BCGs). They are an excellent tool to study the evolution of molecular gas reservoirs in giant ellipticals, even close to the active galactic nucleus.}
    {We investigate the role of dense megaparsec-scale environment in processing molecular gas in LLRGs in the cores of   
galaxy (proto-)clusters. To this aim we selected within the COSMOS and DES surveys a sample of five LLRGs at  $z=0.4-2.6$ that show evidence of ongoing star formation on the basis of their far-infrared (FIR) emission.}
   {We assembled and modeled the FIR-to-UV spectral energy distributions (SEDs) of the five radio sources to characterize their host galaxies in terms of stellar mass and star formation rate. We observed the LLRGs with the IRAM-30m telescope to search for CO emission. We then searched for dense megaparsec-scale overdensities associated with the LLRGs using photometric redshifts of galaxies and the Poisson Probability Method, which we have upgraded using an approach based on the wavelet-transform ($\mathit{w}$PPM), to ultimately characterize the overdensity in the projected space and estimate the radio galaxy miscentering. Color-color and color-magnitude plots were then derived for the fiducial cluster members, selected using photometric redshifts.}
   {Our IRAM-30m observations yielded upper limits to the CO emission of the LLRGs, at $z=0.39, 0.61, 0.91, 0.97$, {and 2.6. For the most distant radio source, COSMOS-FRI~70 at $z=2.6$, a hint of CO(7$\rightarrow$6) emission is found {at 2.2$\sigma$}.} 
   {The upper limits found for} the molecular gas content 
   {$M({\rm H}_2)/M_\star<$0.11, 0.09, 1.8, 1.5, and 0.29, respectively,} and depletion time {$\tau_{\rm dep}\lesssim(0.2-7)$~Gyr} of the five LLRGs are overall consistent with { the corresponding values of main sequence field galaxies.} 
   Our SED modeling implies large stellar-mass estimates in the range $\log(M_\star/M_\odot)=10.9-11.5$, typical for giant ellipticals. Both our $\mathit{w}$PPM analysis and the cross-matching of the LLRGs with existing cluster/group catalogs suggest that the megaparsec-scale overdensities around our LLRGs are rich ($\lesssim10^{14}~M_\odot$) groups  and show a complex morphology. The color-color and color-magnitude plots suggest that the LLRGs are consistent with being star forming and on the high-luminosity tail of the red sequence. The present study thus increases the still limited statistics of distant cluster core galaxies with CO observations.}
   {The radio galaxies of this work are excellent targets for ALMA as well as next-generation telescopes such as the {\it James Webb Space Telescope.}}


\titlerunning{Molecular gas in distant radio galaxies in megaparsec-scale overdensities}
\maketitle
%

\section{Introduction}\label{sec:introduction}
Radio galaxies are powerful extra-galactic sources with low-frequency radio luminosities in the range $\sim10^{41-46}$~erg~s$^{-1}$. They are typically hosted by giant ellipticals and associated with the most massive $\gtrsim10^8~M_\odot$ black holes. The majority of the radio galaxies, at least in the local universe, have little molecular gas to feed star formation and  the central nuclear regions \citep{Lim2000,Lim2004,Evans2005,OcanaFlaquer2010,Baldi2015}.

However in the distant universe ($z\gtrsim2$) large reservoirs of $\gtrsim10^{10}~M_\odot$ of molecular gas have  commonly been found in powerful high-$z$ radio galaxies \citep{Scoville1997,Alloin2000,Papadopoulos2000,DeBreuck2003a,DeBreuck2003b,DeBreuck2005,Greve2004,Klamer2005,Ivison2008,Ivison2011,Nesvadba2009,Emonts2011}. We refer to \citet{Miley_DeBreuck2008} for a review.

Low luminosity radio galaxies (LLRGs) with low-radio frequency luminosities $\lesssim5\times10^{41}$~erg~s$^{-1}$ at 178~MHz represent the great majority among the radio galaxy population, because of the steepness of the radio luminosity function. However, mainly because of their intrinsically low radio power, LLRGs are difficult to find in the distant universe \citep{Snellen_Best2001,Chiaberge2009,Mauch_Sadler2007,Donoso2009,Smolcic2009}

Remarkably, LLRGs are often hosted by giant ellipticals of cD type \citep{Zirbel1996}, which are typically associated with the brightest cluster galaxies \citep[BCGs,][]{von_der_Linden2007,Yu2018}. The vast majority ($\sim70\%$) of LLRGs are in fact found in rich groups and clusters, out to $z\sim2$, almost independently of the redshift \citep{Hill_Lilly1991,Zirbel1997,Wing_Blanton2001,Castignani2014,Castignani2014b,PaternoMahler2017}. LLRGs are therefore a precious tool to search for distant galaxy groups and clusters as well as BCGs.

The BCGs are unique laboratories to study the effect of dense galaxy cluster environment on galaxy evolution.
They are located at the center of the cluster cores \citep{Lauer2014} and exhibit exceptional masses and luminosities. They are believed to evolve via phenomena such as dynamical friction \citep{White1976}, galactic cannibalism \citep{Hausman_Ostriker1978}, interactions with the intracluster medium \citep{Stott2012}, and cooling flows \citep{Salome2006}. In the local universe some studies have shown evidence of molecular gas reservoirs in BCGs \citep{Edge2001,Salome_Combes2003,Hamer2012,McNamara2014,Russell2014,Tremblay2016}, however little is known about the evolution of such reservoirs and the formation of BCGs. 

Recent work suggests that BCGs have doubled their stellar mass since $z\sim1$ \citep{Lidman2012}, which is consistent with a global picture where BCGs evolve via dry accretion of satellite galaxies \citep{Collins2009,Stott2011}. More recent studies have however found potentially conflicting results to this somewhat simplistic hypothesis.  Possible evidence for high levels of star formation and large reservoirs of molecular gas has in fact been suggested for BCGs and cluster core galaxies out to $z\sim1$ and beyond \citep{Brodwin2013,Zeimann2013,Webb2013,Webb2015,Webb2015b,Alberts2016,McDonald2016,Bonaventura2017}, possibly privileging the late assembly of cluster core members via both strong environmental quenching mechanisms 
\citep[e.g., strangulation, ram pressure stripping, and galaxy harassment;][]{Larson1980,Moore1999} and rapid infall of gas feeding star formation at high-$z$ (reaching a maximum at $z\sim2-3$), followed by slow cooling flows at low-$z$ \citep{Ocvirk2008,Dekel2009a,Dekel2009b}.

In this work we study the molecular gas properties of a sample of five star forming LLRGs at $z=0.4-2.6$ that have been selected since they show evidence of significant star formation based on their FIR emission.
With the present work we aim to i) probe the molecular gas content and ii) investigate the role of dense megaparsec-scale environment in processing molecular gas of distant LLRGs in the cores of galaxy (proto-)clusters.
The five radio sources are in fact hosted in dense megaparsec-scale environments and are potentially the high-$z$ progenitors of present day star forming ($>40~M_\odot$/yr) local BCGs such as the famous Perseus~A and Cygnus~A \citep{FraserMcKelvie2014}. 

This work is the second reporting the results of a wider search for molecular gas in distant cluster galaxies \citep[see also][]{Castignani2018}. The paper is structured as follows. In Sect.~\ref{sec:sample} we introduce the LLRG sample; in Sect.~\ref{sec:observations_and_data_reduction} we report our IRAM-30m observations and data reduction; in Sect.~\ref{sec:wPPM} we describe the wavelet-based Poisson Probability Method, which we  developed and applied to search for distant (proto-)clusters around LLRGs; in Sects.~\ref{sec:results} we present our results; in Sect.~\ref{sec:discussion} we discuss the results; and in Sect.~\ref{sec:conclusions} we draw our conclusions.

In this study we refer to megaparsec-scale overdensities, galaxy clusters, galaxy groups, and proto-clusters, with no specific distinction. However we stress that the megaparsec-scale overdensities associated with the LLRGs in our sample  may have different properties. In particular they may be still forming proto-clusters, virialized clusters, or lower-mass groups \citep[see e.g.,][for a review]{Overzier2016}.

Throughout this work we adopt a flat $\Lambda \rm CDM$ cosmology with matter density $\Omega_{\rm m} = 0.30$, dark energy density $\Omega_{\Lambda} = 0.70,$ and Hubble constant $h=H_0/100\, \rm km\,s^{-1}\,Mpc^{-1} = 0.70$ \citep[see however,][]{Planck2018resultsVI,Riess2016,Riess2018}.

\section{The radio galaxy sample}\label{sec:sample}
The Dark Energy Survey \citep[DES,][]{des2005,des2016} is an ongoing five-year project (2013-2018) composed of two distinct multi-band imaging surveys: a $\sim$5000~deg$^2$ wide-area \textsf{grizY} survey and a deep supernova \textsf{griz} survey made by six distinct deep fields \citep{Kessler2015}. The coadded source catalog and images from the first 3~yr of science operations have recently been made public as part of the first public data release \citep[DES DR1,][]{Abbott2018}\footnote{https://des.ncsa.illinois.edu/releases/dr1}.

The Very Large Array Faint Images of the Radio Sky at Twenty-centimeters (VLA FIRST) survey  \citep{Becker1995} observed 10,000~deg$^2$ of the North and South Galactic Caps at 1.4~GHz. Post-pipeline radio maps have a typical full width at the half maximum (FWHM) resolution of $\sim$5~arcsec. The detection limit of the FIRST source catalog is $\sim$1~mJy with a typical rms of 0.15~mJy.

In order to select the radio galaxy sample we have limited ourselves to the DES supernova (SN) deep fields that overlap with the FIRST survey. This selection yielded four fields, namely the DES SN deep fields numbered 2, 3, 5, and 6. Field number 6 is located at approximately (R.A.~;~Dec.)=(209.5~;~4.9)~deg and covers $\sim4$~deg$^2$. It is not included in the DES DR1 survey area, and is therefore not considered in this work. Fields 2 and 3 are located at approximately (R.A.~;~Dec.)~=~(35.5~;~-5.5)~deg and (42.0~;~-0.4)~deg, and subtend $\sim$16~deg$^2$ and $\sim$10~deg$^2$ of solid angle in the sky, respectively. They are both included in the DES DR1 and are therefore considered in this work. Field number 5 is located at approximately (R.A.~;~Dec.)=(150.0~;~2.2)~deg and covers $\sim4$~deg$^2$. It is not included in the DES DR1, but it entirely includes the 2~deg$^2$ Cosmic Evolution Survey \citep[COSMOS,][]{Scoville2007} that is considered in this work.  Concerning the sample selection and characterization in the following we separately consider the DES SN deep fields 2 and 3 (Sect.~\ref{sec:DES_RGs}), and the COSMOS survey (Sect.~\ref{sec:COSMOS_FRIs}).

{ With the present work we aim at studying the molecular gas properties of a pilot sample of rare distant BCG candidates with suggested evidence of ongoing star formation. For this pilot study we considered the DES SN deep fields because they partially overlap with the VLA FIRST survey and because their deep observations (down to AB magnitudes $\sim$24.5 in all DES bands) enable us to effectively find and characterize both distant LLRGs and their megaparsec-scale environments. }

\subsection{DES radio galaxies 399 and 708}\label{sec:DES_RGs}
We aim at selecting radio galaxies to be followed-up with radio facilities at millimeter (mm) wavelengths. Accurate redshift measurements are therefore needed: they have to be used as a positional prior for the follow-ups. We consider DES DR1 sources in the DES SN deep fields 2 and 3 and search for their spectroscopic redshifts from both the Fourteenth Data Release\footnote{http://www.sdss.org/dr14/} (DR14) of the Sloan Digital Sky Survey (SDSS) and the DES spectroscopic dataset (C.~Benoist, private comm.). Since the astrometric precision of SDSS and DES is $\simeq0.1$~arcsec we adopted a search radius of 1~arcsec. This selection yielded 33,148 unique spectroscopic counterparts. 
We aim at studying distant radio galaxies. We therefore limit ourselves to galaxies with spectroscopic redshifts $z>0.3$. This selection yielded 22,778 spectroscopic sources.

We have adopted the DES magnitudes corrected for Galactic extinction, as listed in the DES DR1 coadd catalog (DR1$\_$MAIN) and denoted, for example, \textsf{MAG$\_$AUTO$\_$G$\_$DERED}  \citep[we refer to Appendix~D of][for further details]{Abbott2018}.

\subsubsection{FIRST}
We searched for radio counterparts of the DES sources using the FIRST source catalog and a search radius of 3~arcsec, consistently with the positional accuracy $\sim1$~arcsec of FIRST sources. The search yielded 151 unique low-frequency radio counterparts.

\subsubsection{WISE}\label{sec:WISE}
One of the main goals of this work is to select a sample of star forming radio galaxies. We have therefore selected a sub-sample of radio sources with 22~$\mu$m emission in the observer frame, as found by the W4 filter of the Wide-field Infrared Survey Explorer  \citep[WISE,][]{Wright2010}. We have cross-correlated our DES sources with the allWISE source catalog\footnote{http://wise2.ipac.caltech.edu/docs/release/allwise/} by adopting a search radius of 6.5~arcsec, consistently with previous studies on distant radio sources \citep[e.g.,][]{Castignani2013}. The search yielded 50 sources with  unique WISE counterparts and W4 magnitudes with signal-to-noise ratio S/N $>1$. A detailed study of such a sample will be performed in a forthcoming paper (Castignani et al., in prep.). \\ 

Among the 50 sources we have selected two galaxies that we observed with the IRAM-30m telescope, as part of a pilot project to search for CO in distant star-forming galaxies. The two selected galaxies are both within the DES SN deep field number 3 and are denoted hereafter as DES-RG~399 and 708, where RG stands for radio galaxy. {They have WISE counterparts in the allWISE source catalog with W4 magnitudes} { reported at 2.1$\sigma$ and 1.5$\sigma,$} respectively. {In the other WISE  W1, W2, and W3 channels the sources are detected at ${\rm S/N}>9.5$, with the exception of DES-RG~708 whose WISE W3 emission at 12~$\mu$m in the observer frame is found { at 2.5$\sigma$.}} 

The two galaxies have spectroscopic redshifts $z=0.39$ and $0.61$, respectively. In order to characterize the broad band emission of both sources we searched for additional data using the DES coordinates as input, as described in the following.

\subsubsection{NVSS}\label{sec:NVSS}
The NRAO VLA Sky Survey (NVSS) survey \citep{Condon1998} at 1.4~GHz was obtained using the
VLA-D configuration. The FWHM angular resolution of the NVSS radio maps is 45~arcsec. Therefore, NVSS is more suitable than FIRST for detecting extended emission. We found NVSS counterparts for both DES-RG~399 and 708, at an angular separation of 1.2~arcsec and 2.4~arcsec, respectively. Their NVSS fluxes are (12.0$\pm$0.6)~mJy and (36.0$\pm$1.5)~mJy, respectively, while the FIRST fluxes are 12.54~mJy and 11.80~mJy. An additional southern source is detected by FIRST at an angular separation of $\sim$7.2~arcsec from DES-RG~708 with a flux of 12.63~mJy; it is visible in the VLA-FIRST image (Fig.~\ref{fig:DES_RG_images}, top).
DES-RG~708 might have an extended radio morphology, unresolved by NVSS. Alternatively, it is also possible that the radio source has a radio morphology with two hot-spots, typical of FR~II radio sources \citep{Fanaroff_Riley1974}.
The available angular resolution does not allow us to distinguish between the two scenarios. 

\subsubsection{IRAS}
We searched for far-infrared (FIR) counterparts of our sources within the Infrared Astronomical Satellite (IRAS) Faint Source Catalog \citep{Moshir1990} and the IRAS Point Source Catalog \citep{Helou_Walker1988}.  Neither of the two DES-RGs was found.
We extracted 15$'\times$15$'$ IRAS Sky Survey Atlas (ISSA) image cutouts\footnote{https://irsa.ipac.caltech.edu/data/ISSA/index$\_$cutouts.html} 
at 12, 25, 60, and 100~$\mu$m, centered at the coordinates of our DES-RG sources. The pixel size of the images is 90~arcsec that corresponds to $\sim$(500-600)~kpc for DES-RG~399 and 708, which are therefore unresolved. We then estimated { 3$\sigma$} upper limits to the IRAS fluxes as three times the rms dispersion derived from the cutout images  of the two sources.

\subsubsection{UKIDSS}\label{sec:UKIDSS}
The UKIRT Infrared Deep Sky Survey (UKIDSS) mapped the near-infrared (NIR) sky \citep{Lawrence2007}. UKIDSS used the UKIRT Wide Field Camera \citep[WFCAM,][]{Casali2007}. The photometric system is described in \citet{Hewett2006}, and the calibration is described in \citet{Hodgkin2009}. The pipeline processing and science archive are described in \citet{Hambly2008}.
The UKIDSS Large Area Survey (LAS)\footnote{http://www.ukidss.org/surveys/surveys.html} consists of observations covering 7,500~deg$^2$ of the sky at the \textsf{YJHK} NIR bands and includes the DES SN deep field number 3. We searched within the latest public release UKIDSSDR10PLUS and found unique UKIDSS counterparts for both  DES sources using a search radius of 3~arcsec, which is consistent with the positional FWHM accuracy $\lesssim1.2$~arcsec of UKIDSS \citep{Lawrence2007}. The two counterparts are located at angular separations of 0.5~arcsec and 0.2~arcsec from the DES coordinates of DES-RG~399 and 708, respectively. UKIDSS Vega magnitudes, derived within a 2-arcsec aperture, were converted into the AB system by applying the offsets reported in Table~7 of \citet{Hewett2006}. The magnitudes were also corrected for Galactic extinction using  the extinction curve by \citet{Cardelli1989}, as updated by \citet{ODonnell1994} and normalized to ${\rm A(V)} = 3.1\,{\rm  E(B-V)}$. For each source, the value of ${\rm E(B-V)}$ was estimated using the extinction values reported for the SDSS bands and the prescriptions described in SDSS DR14 tutorial\footnote{http://www.sdss.org/dr14/spectro/sspp/$\#$ExtinctionCalculations}.

\subsubsection{SDSS}
The DES SN deep field number 3 is entirely contained within the SDSS survey area. We found unique SDSS DR14 optical counterparts at the \textsf{u}, \textsf{g}, \textsf{r}, \textsf{i}, and \textsf{z} bands of SDSS for both sources using a search radius of 1~arcsec.  We adopted the SDSS magnitudes corrected for Galactic extinction, as listed in the DR14 catalog and denoted, for example, \textsf{dered$\_$g}. As suggested in the DR14 tutorial\footnote{http://www.sdss.org/dr14/algorithms/fluxcal/} we decreased the DR14 \textsf{u}-band magnitudes by 0.04 to convert them into the AB system. For the other bands the corrections are negligible.
In Fig.~\ref{fig:DES_RG_images} (bottom) we show the SDSS images of DES-RG~399 and 708.

\subsubsection{GALEX}
\begin{figure*}[t]\centering
\subfloat{\includegraphics[width=0.3\textwidth]{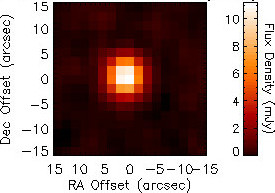}}
\subfloat{\hspace{2cm}\includegraphics[width=0.3\textwidth]{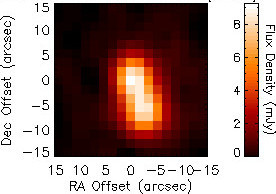}}\\
\subfloat{\includegraphics[width=0.165\textwidth]{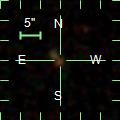}}
\hspace{4.7cm}\subfloat{\includegraphics[width=0.165\textwidth]{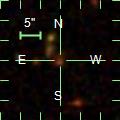}}
\caption{VLA FIRST (top) and composite SDSS DR14 (bottom) images centered at the DES coordinates of DES-RG~399 (left) and DES-RG~708 (right). All images have 30$''\times$30$''$ sizes and north is up, east is left.   }
\label{fig:DES_RG_images}
\subfloat{\includegraphics[width=0.3\textwidth]{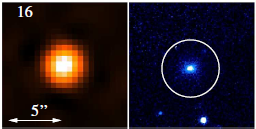}}
\subfloat{\hspace{0.5cm}\includegraphics[width=0.3\textwidth]{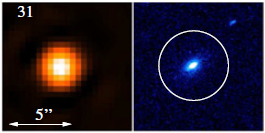}}
\subfloat{\hspace{0.5cm}\includegraphics[width=0.3\textwidth]{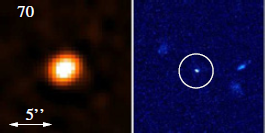}}\\
\caption{Radio and optical images from \citet{Chiaberge2009} for the COSMOS-FRI sources in our sample. For each source, the
image in the left panel is from the VLA-COSMOS survey \citep{Schinnerer2007}, while in the right panel the HST-COSMOS ACS image (F814W) is shown \citep{Koekemoer2007}.}
\label{fig:COSMOS_FRI_images}
\end{figure*}

We searched for UV photometry of our DES radio sources in the joint sixth and seventh data releases, GR~6/7,\footnote{http://galex.stsci.edu/GR6/} of the Galaxy Evolution Explorer (GALEX) satellite \citep{Morrissey2007}. GALEX provides near-UV (NUV, 1750-2800$\AA$) and far-UV (FUV, 1350-1750$\AA$) source fluxes down to a magnitude limit AB$\sim$20-21 with an estimated positional accuracy of $\sim$0.5 arcsec. 
By adopting a search radius of 2~arcsec we found a possible GALEX counterpart for one of the two sources, namely DES-RG~399, at an angular separation of 1.8~arcsec.
However DES-RG~399 has a companion located at 1.8~arcsec of angular separation, visible in the SDSS image (see Fig.~\ref{fig:DES_RG_images}).
The GALEX source is located at an angular separation of 1.3~arcsec from this companion and it is therefore possible that it is not associated with the source DES-RG~399. Visual inspection of the IR-to-UV spectral energy distribution (SED) of DES-RG~399 further strengthens such a possibility. We therefore preferred to remove the GALEX association for DES-RG~399.

In Table~\ref{tab:DES_RG_photometry} we summarize the IR-to-optical photometric data of the two DES-RG sources. 
{ We note that DES and SDSS magnitudes of the two DES-RG sources are discrepant up to $\sim0.7$~mag. DES-RG~399 and 708 are primarily selected at low radio frequency (FIRST) by adopting a search radius of 1~arcsec that is used to look for the optical DES counterparts. This radio positional uncertainty implies that we cannot completely rule out the possibility that the discrepancy in the magnitudes is due to the companions of the radio sources, which are in fact visible in the optical images at angular separations of the order of $\sim1$~arcsec (see Fig.~\ref{fig:DES_RG_images}, bottom).} 

\subsection{COSMOS-FRI radio galaxies 16, 31, and 70}\label{sec:COSMOS_FRIs}
\subsubsection{Target selection}
COSMOS is a multiwavelength equatorial 2~deg$^2$  survey \citep[][]{Scoville2007} that includes Hubble Space Telescope (HST) observations \citep{Koekemoer2007} and IR Spitzer imaging \citep[S-COSMOS,][]{Sanders2007}. We limit ourselves to the sample of well-studied distant $z\sim1-3$ FR~I \citep{Fanaroff_Riley1974} radio galaxy candidates  described in \citet{Chiaberge2009} and further reconsidered on the basis of their radio power by \citet{Castignani2014}.

With the aim of selecting a sample of distant radio galaxies to be followed-up with radio facilities at mm wavelengths we restrict ourselves to the subsample of COSMOS-FRIs with available spectroscopic redshifts \citep[see ][for further details]{Baldi2013,Castignani2014}, similarly to what was done in Sect.~\ref{sec:DES_RGs}. This selection yielded eight sources, namely, COSMOS-FRI~1, 16, 27, 66, 31, 52, 70, and 258.
A ninth source, 236, is included in the \citet{Chiaberge2009} sample and has a spectroscopic redshift. However it is excluded from our analysis because it is a confirmed QSO at $z=2.132$ \citep{Prescott2006}. The source COSMOS-FRI~70 has no spectroscopic redshift from the literature. However we verified (M. Bolzonella, private comm.) that it is the only COSMOS-FRI with a spectroscopic counterpart within the zCOSMOS-deep catalog \citep[][Lilly et al., in prep.]{Lilly2007} \footnote{http://archive.eso.org/cms/eso-data/data-packages/zcosmos-data-release-dr1.html}, which includes approximately 10,000 galaxies of the COSMOS survey
at approximately $1.5<z<3.0$. We therefore included COSMOS-FRI~70 in our analysis. 

\citet{Baldi2013}, hereafter B13, carefully reconsidered the IR, optical, and UV photometric data associated with the COSMOS-FRIs of the \citet{Chiaberge2009} sample and performed SED fits to the data. On the other hand the associations provided in the COSMOS photometric catalogs are performed automatically \citep[e.g.,][]{Ilbert2009,Laigle2016}.
Therefore in this work we adopt the photometric data provided in B13 for the COSMOS-FRI galaxies.
In particular, since we aim to select star forming galaxies we restrict ourselves to the subsample of sources with available 
Spitzer MIPS flux at 23.68$\mu$m from B13.
This selection yielded three sources, namely COSMOS-FRI~16, 52, and 70. We also included  COSMOS-FRI~31 because it has a
WISE counterpart with W4 magnitude reported at { 0.9$\sigma$} in the allWISE source catalog, while we discarded COSMOS-FRI~52 because we verified that its redshifted CO lines cannot be optimally observed. Therefore we limit our analysis to COSMOS-FRI~16, 31, 70.  In Fig.~\ref{fig:COSMOS_FRI_images} we report the radio and optical images of the three COSMOS-FRI sources considered.

\begin{table*}[htb]\centering
\begin{tabular}{cccc}
\hline\hline
 survey & band  & DES-RG~399  & DES-RG~708 \\
 (1)    & (2)   & (3)         & (4) \\
 \hline\hline
SDSS & \textsf{u}  & 22.14$\pm$0.58 & 23.87$\pm$1.17 \\
     & \textsf{g}  & 21.09$\pm$0.09 & 22.48$\pm$0.17 \\
     & \textsf{r}  & 20.45$\pm$0.08 & 21.22$\pm$0.08 \\
     & \textsf{i}  & 19.64$\pm$0.06 & 20.19$\pm$0.05 \\
     & \textsf{z}  & 19.11$\pm$0.14 & 19.70$\pm$0.12 \\     
\hline
DES  & \textsf{g}  & 21.74$\pm$0.04 & 22.10$\pm$0.06 \\
     & \textsf{r}  & 20.79$\pm$0.02 & 20.55$\pm$0.02 \\
     & \textsf{i}  & 20.13$\pm$0.02 & 19.52$\pm$0.01 \\
     & \textsf{z}  & 19.79$\pm$0.02 & 19.08$\pm$0.02 \\
     & \textsf{Y}  & 19.62$\pm$0.07 & 18.82$\pm$0.04 \\
\hline
UKIDSS  & \textsf{Y}  & 19.33$\pm$0.10 & 19.12$\pm$0.08 \\
        & \textsf{J}  & 18.81$\pm$0.11 & 18.63$\pm$0.09 \\
        & \textsf{H}  & 17.71$\pm$0.08 & 17.65$\pm$0.07 \\
        & \textsf{K}  & 16.64$\pm$0.04 & 17.30$\pm$0.08 \\
\hline
WISE    & W1  & 15.213$\pm$0.032 & 14.581$\pm$0.030\\
        & W2  & 14.929$\pm$0.069 & 14.392$\pm$0.057 \\
        & W3  & 11.124$\pm$0.114 & 12.324$\pm$0.426 \\
        & W4  & 9.090$\pm$0.506 & 8.403$\pm$0.724 \\
\hline
IRAS    & ${12\mu {\rm m}}$ &  $<$30.06~mJy & $<$58.66~mJy \\
        & ${25\mu {\rm m}}$ &  $<$35.71~mJy & $<$93.34~mJy\\
        & ${60\mu {\rm m}}$ &  $<$55.22~mJy & $<$57.42~mJy \\
        & ${100\mu {\rm m}}$ & $<$84.64~mJy & $<$90.94~mJy \\
\hline
\end{tabular}
\caption{Photometry of DES-RG~399 and 708: (1) survey; (2) reference band; (3-4) photometric data associated with the sources, along with their uncertainties. The reported SDSS (AB) magnitudes are \textsf{dered$\_$u}, \textsf{g}, \textsf{r}, \textsf{i}, and \textsf{z}; DES (AB) magnitudes are \textsf{MAG$\_$AUTO$\_$G, R, I, Z, and Y$\_$DERED}; UKIDSS magnitudes are estimated within 2~arcsec aperture and are in the Vega system; WISE W1, W2, W3, and W4 magnitudes are observed at 3.4, 4.6, 12, 22~$\mu$m and are in the Vega system; IRAS fluxes are { 3$\sigma$} upper limits. See text for further details. }
\label{tab:DES_RG_photometry}
\end{table*}
\begin{table*}[hbt]
\begin{center}
\begin{tabular}{ccccc}
\hline\hline
 telescope/survey & band  & COSMOS-FRI~16  & COSMOS-FRI~31 & COSMOS-FRI~70 \\
 (1)    & (2)   & (3)         & (4) & (5) \\
 \hline\hline
CFHT  & \textsf{u$^\ast$}  & 26.07$\pm$0.12  & 24.75$\pm$0.08  & 26.16$\pm$0.14 \\ 
      & \textsf{i$^\ast$}   & 23.10$\pm$0.14   & 22.33$\pm$0.08  & 24.34$\pm$0.58 \\
      & \textsf{K}   & 20.50$\pm$0.11 & 20.32$\pm$0.09  & 21.40$\pm$0.27 \\  
\hline
Subaru& \textsf{B$_{\rm J}$}    & 25.71$\pm$0.12 & 24.50$\pm$0.06  & 25.14$\pm$0.09 \\ 
      & \textsf{g$^+$}    & 25.32$\pm$0.11 & 24.51$\pm$0.05  & 25.27$\pm$0.14 \\ 
      & \textsf{V$_{\rm J}$}    & 24.84$\pm$0.08 & 23.94$\pm$0.04  & 24.79$\pm$0.08 \\        
      & \textsf{r$^+$}    & 24.23$\pm$0.05 & 23.42$\pm$0.03 & 24.51$\pm$0.06 \\
      & \textsf{i$^+$}   & 23.15$\pm$0.03 & 22.33$\pm$0.02  & 24.35$\pm$0.06 \\
      & \textsf{z$^+$}    & 22.16$\pm$0.02 & 21.61$\pm$0.02  & 24.01$\pm$0.09 \\
\hline
HST/ACS & F814W   & 22.85$\pm$0.10 & 22.07$\pm$0.07  & 24.13$\pm$0.43 \\
\hline
UKIRT & \textsf{J}   & 21.47$\pm$0.08 & 20.97$\pm$0.08  & 23.42$\pm$0.50 \\ 
\hline
NOAO  & \textsf{K$_{\rm s}$}   & 20.40$\pm$0.08  & 20.12$\pm$0.05  & 21.76$\pm$0.09 \\
\hline     
GALEX & FUV   & ---  & ---  &  --- \\
      & NUV    & ---  & 25.15$\pm$0.24   & --- \\
\hline
Spitzer  & IRAC1    & (44.29$\pm$0.18)~$\mu$Jy & (47.86$\pm$0.16)~$\mu$Jy  & (17.58$\pm$0.16)~$\mu$Jy \\             
         & IRAC2    & (32.54$\pm$0.28)~$\mu$Jy & (34.60$\pm$0.25)~$\mu$Jy  & (20.57$\pm$0.26)~$\mu$Jy \\
         & IRAC3    & (26.19$\pm$0.94)~$\mu$Jy & (24.98$\pm$0.81)~$\mu$Jy & (29.83$\pm$1.04)~$\mu$Jy \\
         & IRAC4    & (18.39$\pm$2.26)~$\mu$Jy & (19.36$\pm$1.78)~$\mu$Jy & (21.32$\pm$2.17)~$\mu$Jy \\
         & MIPS (24$\mu$m) & (0.22$\pm$0.02)~mJy &  $<0.15$~mJy & (0.13$\pm$0.03)~mJy \\
         & MIPS (70$\mu$m) & $<5.1$~mJy &  $<5.1$~mJy & $<5.1$~mJy \\
         & MIPS (160$\mu$m)& $<39$~mJy &  $<39$~mJy & $<39$~mJy \\
\hline
WISE     & W1    & 16.923$\pm$0.114 & 17.360$\pm$0.162  & >17.672 \\    
         & W2    & 16.711$\pm$0.388 & 16.924$\pm$0.418  & >18.324 \\
         & W3    & >10.680  &  >16.955 & >16.955 \\
         & W4    & >7.712   & >7.283  & >15.009 \\
\hline
\end{tabular}
\end{center}
\caption{Photometry of COSMOS-FRI~16, 31, and 70: (1) survey/telescope; (2) reference band; (3-5) photometric data associated with the sources, along with their uncertainties. Magnitudes from CFHT, Subaru, HST/ACS, UKIRT, NOAO, and GALEX are in AB system and from B13. Spitzer IRAC and MIPS (24$\mu$m) fluxes are also from B13.
We report the symbol $''$---$''$ when the GALEX magnitudes are absent. 
WISE W1, W2, W3, and W4 magnitudes and magnitude lower limits are in the Vega system.}
\label{tab:COSMOS_FRI_photometry}
\vspace*{1cm}
\end{table*}

\subsubsection{Infrared-to-ultraviolet photometric data} 
For the three COSMOS sources the IR-to-optical photometry reported by B13 and considered in this work includes data from the Canada-France-Hawaii Telescope \citep[CFHT,][]{Boulade2003}, UKIRT \citep{Casali2007}, NOAO \citep{Capak2007}, Subaru \citep{Taniguchi2007}, Spitzer IRAC and Spitzer MIPS at 23.68$\mu$m  \citep{Sanders2007}, HST \citep{Koekemoer2007}, and GALEX \citep{Morrissey2007}. We also included { 3$\sigma$} upper limits for the Spitzer MIPS fluxes at 70$\mu$m and 160$\mu$m, corresponding to 5.1~mJy and 39~mJy, respectively \citep[][]{Lee2010}. 

Following a procedure similar to that described in Sect.~\ref{sec:WISE}, WISE counterparts were found for all three sources except COSMOS-FRI~70. {Similarly, COSMOS-FRI~31 is not detected in the WISE W3 channel.} Analogously to previous work by \citet{Castignani_DeZotti2015}, in these cases, where WISE fluxes are absent, we have adopted 3$\sigma$  flux density upper limits of 0.6 and 3.6~mJy, due to instrumental noise alone \citep{Wright2010}, for the WISE channels W3 and W4, respectively. For channels W1 and W2 the 3$\sigma$ limits are set by confusion noise and are equal to 0.31 and 0.17~mJy, respectively \citep{Jarret2011}. { COSMOS-FRI~31, in the W4 channel, and COSMOS-FRI~16, in both W3 and W4 channels, are only marginally detected by WISE, at S/N<2. We therefore converted the associated fluxes into {3$\sigma$} upper limits.}
{ In Table~\ref{tab:COSMOS_FRI_photometry} we summarize the IR-to-UV photometry of COSMOS-FRI~16, 31, and 70. }

The NIR-to-optical magnitudes of these three COSMOS-FRI sources were then corrected for Galactic extinction by adopting a procedure analogous to that described in Sect.~\ref{sec:UKIDSS}. For each source the value of ${\rm E(B-V)}$ was obtained assuming the value for Galactic extinction estimated by \citet{Schlafly_Finkbeiner2011} at the source location and normalized to ${\rm A(V)} = 3.1\,{\rm  E(B-V)}$.

Our final sample comprises five sources, DES-RG~399, DES-RG~708, COSMOS-FRI~16, COSMOS-FRI~31, and COSMOS-FRI~70. { In the following we describe the SED modeling for these sources.}

\begin{figure*}[h!]
\begin{center}
\subfloat{\includegraphics[width=0.3\textwidth]{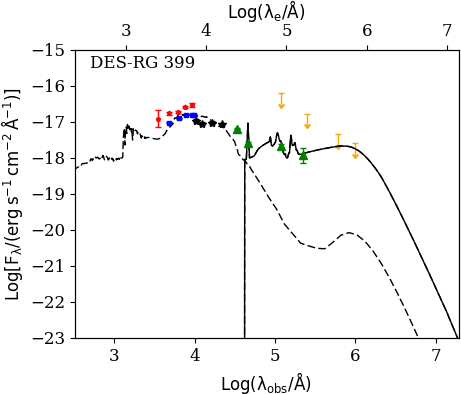}}
\subfloat{\hspace{0.5cm}\includegraphics[width=0.3\textwidth]{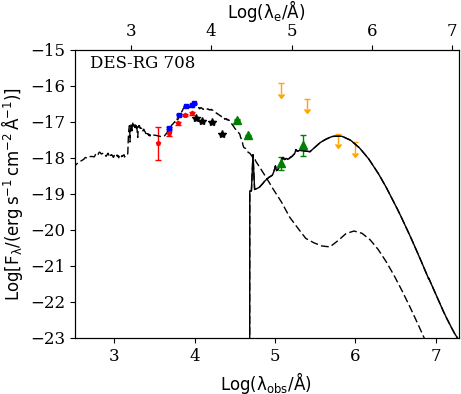}}\\
\subfloat{\hspace{0.5cm}\includegraphics[width=0.3\textwidth]{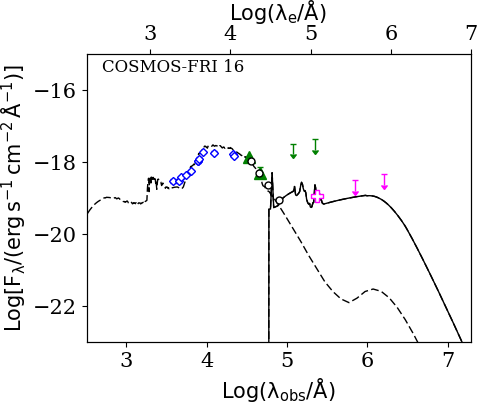}}
\subfloat{\hspace{0.5cm}\includegraphics[width=0.3\textwidth]{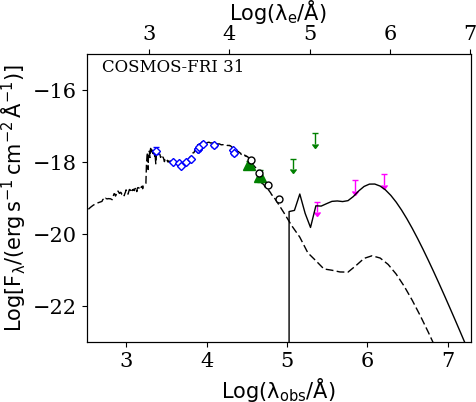}}
\subfloat{\hspace{0.5cm}\includegraphics[width=0.3\textwidth]{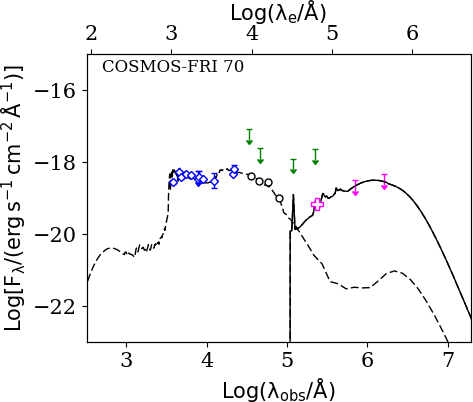}}\\
\end{center}
\caption{SEDs and modeling for DES-RG (top) and COSMOS-FRI (bottom) sources in the sample. Data-points for DES-RG sources are from SDSS (red pentagons), DES (blue squares), UKIDSS (black stars), WISE (green triangles), and IRAS (yellow upper limits), while those for COSMOS-FRI galaxies are archival NIR-to-UV data from B13 (blue open diamonds). Spitzer-IRAC (open black dots), Spitzer-MIPS (pink open crosses and upper limits), and WISE (green open triangles and upper limits) data are also included for COSMOS-FRI galaxies. See text for further details. Dashed and solid lines show the best fit models for the stellar and dust components, respectively.}
\label{fig:SEDs}
\end{figure*}

\subsection{SED modeling}\label{sec:SEDmodeling}
{We assembled the IR-to-UV SEDs for the five sources in our sample using the photometric dataset described above and performed fits to the SEDs using LePhare \citep{Arnouts1999,Ilbert2006}.
Following the prescriptions provided for the LePhare code, we fitted the FIR data separately to account for possible dust emission, using the  \citet{Chary_Elbaz2001} library consisting of 105 templates.
The remaining photometric points at shorter wavelengths were fitted using the CE$\_$NEW$\_$MOD library provided by LePhare, which is similar to that described in \citet{Arnouts1999} and consists of 66 templates based on linear interpolation of the four original SEDs of \citet{Coleman1980}. The SEDs along with their best fits are reported in Fig.~\ref{fig:SEDs} for all five sources in our sample.

We stress that B13 already reported the IR-to-UV SED modeling for the COSMOS-FRI sources considered in this work. However we preferred to perform the SED fitting independently because we want to report homogeneous results for all galaxies considered in this work, including the DES-RG sources not considered by B13. Furthermore, B13 did not include Spitzer MIPS upper limits at 70 and 160$\mu$m and did not provide SFR estimates; see also the following sections. }

\subsection{Radio luminosities}
{In this section we estimate the radio luminosities for the five radio sources in our sample. }
Similarly to previous work \citep{Chiaberge2009,Castignani2014} we assume that the radio spectrum in the region around the frequency 1.4~GHz is a power-law of the form $S_\nu\propto\nu^{-\alpha}$,  where $S_\nu$ is the radio flux density at the observer frequency $\nu$ and the spectral index $\alpha$ is assumed to be $\alpha=0.8$.
We adopt the NVSS flux densities reported in Sect.~\ref{sec:NVSS} for DES-RG~399 and 708, while those of COSMOS-FRI~16, 31, and 70 are taken from \citet{Castignani2014}. We then estimate the rest frame 1.4~GHz luminosity density as follows:
\begin{equation}
\label{eq:L14formula} 
L_{1.4~{\rm GHz}} = 4\pi S_{1.4~{\rm GHz}}D_L(z)^2\left(1+z\right)^{\alpha-1},
\end{equation}
where $S_{1.4~{\rm GHz}}$ is the observed 1.4~GHz flux density, D$_L$ is the luminosity distance, and $\alpha$ is the radio spectral index. Low-radio frequency and additional properties for all five sources in our sample are summarized in Table~\ref{tab:radio_galaxies_properties0}.

\begin{table*}[htb]
\begin{adjustwidth}{-0.7cm}{}
\begin{tabular}{ccccccccc}
\hline\hline
Galaxy ID & R.A. & Dec. & $z_{spec}$ & $S_{\rm 1.4~GHz}^{\rm FIRST}$ & $S_{\rm 1.4~GHz}^{\rm NVSS}$ & $L_{\rm 1.4~GHz}$ & {\small Radio } & { \small Optical} \\
   & (hh:mm:ss.sss) & (dd:mm:ss.sss) & & (mJy) & (mJy) & {\small (10$^{32}$~erg~s$^{-1}$~Hz$^{-1}$)}  & Morphology &   Morphology \\ 
 (1) & (2) & (3) & (4) & (5) & (6) & (7) & (8) & (9) \\
 \hline
 {\small DES-RG~399} & {\small 02:42:27.221} & {\small -00:34:41.732}  & {\small 0.38844$\pm$0.00004} & 12.54 & 12.0 & 0.59 & unresolved & smooth \\
 \hline
 {\small DES-RG~708} & {\small 02:45:21.751} & {\small -00:32:29.782}  & {\small 0.60573} & 11.80 & 36.0 & 5.00 & extended & smooth \\
 \hline 
 {\small COSMOS-FRI~16} & {\small 10:02:09.053} & {\small +02:16:02.478}  & {\small 0.9687$\pm$0.0005} & 5.70 & 4.4 & 1.86 & unresolved & smooth \\
 \hline
 {\small COSMOS-FRI~31} & {\small 09:58:28.598} & {\small +01:54:58.896}  & {\small 0.9123$\pm$0.0003} & 3.71 & 4.1 & 1.50 & compact & smooth \\
 \hline
 {\small COSMOS-FRI~70} & {\small 10:02:28.769}  & {\small +02:17:21.970}  & {\small 2.625$\pm$0.003} & 3.90 & 4.5 & 19.49 & compact & --- \\
 \hline
\end{tabular}
\caption{Properties of our targets: (1) galaxy name; (2-3) J2000 equatorial coordinates, for the sources DES-RG~399 and 708 the coordinates of the DES counterparts are reported, while those of COSMOS-FRI~16, 31, and 70 are from \citet{Castignani2014}; (4) spectroscopic redshift and uncertainty of DES-RG~399 \citep[from the BOSS survey,][as found in the DR14 of SDSS]{Dawson2013}, of DES-RG~708 (DES spectroscopic dataset; C.~Benoist, private comm.), of COSMOS-FRI~16 and 31 \citep[from  MAGELLAN,][]{Trump2007}, and of COSMOS-FRI~70 (zCOSMOS-deep; M.~Bolzonella, private comm.); (5-6) FIRST and NVSS flux densities; (7) rest frame 1.4~GHz luminosity density estimated using Eq.~\ref{eq:L14formula}; (8-9) radio and optical morphology inferred from the images reported in Figs.~\ref{fig:DES_RG_images} and \ref{fig:COSMOS_FRI_images}, the morphology of COSMOS-FRI sources is from \citet{Chiaberge2009}.}
\label{tab:radio_galaxies_properties0}
\end{adjustwidth}
\end{table*}

The radio power of our sources is fairly consistent with that of low-luminosity FR~I radio galaxies belonging to the FR~I class, for which $L_{\rm 1.4~GHz}\lesssim2.6\times10^{32}$~erg~s$^{-1}$~Hz$^{-1}$.\footnote{\citet{Fanaroff_Riley1974} originally reported an FR~I/FR~II radio-power  divide of $\sim2\times10^{32}$~erg~s$^{-1}$~Hz$^{-1}$~sr$^{-1}$ at 178~MHz, which we have converted to 1.4~GHz in the rest frame by assuming an isotropic emission and a radio spectral slope $\alpha=0.8$, as well as  accounting for the different cosmology adopted by the authors.} 
From the values reported in Table~\ref{tab:radio_galaxies_properties0} we observe that only DES-RG~708 and COSMOS-FRI~70 have values of $L_{\rm 1.4~GHz}$ that are higher than the reported FR~I/FR~II radio-power divide. However we stress that their radio luminosities are still fairly consistent with the high-power tail of FR~I radio galaxies, since they are at least one order of magnitude less bright than powerful distant radio galaxies at similar redshifts \citep{Miley_DeBreuck2008,Ineson2013}. We refer to \citet{Castignani2014} for further discussion about the radio power properties of the COSMOS-FRI sources.

In the following we describe our IRAM-30m observations and data analysis.

\section{IRAM-30m observations and data reduction}\label{sec:observations_and_data_reduction}
We observed the five radio galaxies in our sample using the IRAM-30m telescope at Pico Veleta in Spain. The observations of our targets were carried out in the summers of 2016 and 2017 as part of two observational programs (P.I.: Castignani).

We used the Eight Mixer Receiver (EMIR) and its E230 band to observe a CO(J$\rightarrow$J-1) emission line from each target source, at frequencies between 215 and 249~GHz, where J is a positive integer denoting the total angular momentum. For each source the specific CO(J$\rightarrow$J-1) transition was chosen to maximize the likelihood for the detection in terms of the ratio of the predicted signal to the expected rms noise. We refer to Table~\ref{tab:radio_galaxies_properties_mol_gas} for further details.

\begin{table*}[htb]\centering
\vspace*{1cm}
\begin{adjustwidth}{-0cm}{}
\begin{tabular}{ccccccccccccc}
\hline\hline
 Galaxy ID &  $z_{spec}$ & CO(J$\rightarrow$J-1)  & $\nu_{\rm obs}$ & $S_{\rm CO(J\rightarrow J-1)}$   &  $M({\rm H_2})$ & $\tau_{\rm dep}$ & $\frac{M({\rm H_2})}{M_\star}$  & $\tau_{\rm dep, MS}$ & $\big(\frac{M({\rm H_2})}{M_\star}\big)_{\rm MS}$  \\
   &  & & (GHz) &  (Jy~km~s$^{-1}$)  & ($10^{10}~M_\odot$) & ($10^9$~yr) &  & ($10^9$~yr) & \\ 
 (1) & (2) & (3) & (4) & (5) & (6) & (7) & (8) & (9) & (10)  \\
 \hline
{\small DES-RG~399} &    {\small 0.388439} & 3$\rightarrow$2 & 249.054  & $<1.5$ & $<$1.0 & $<$0.48 & $<$0.11  &  $1.10^{+0.16}_{-0.14}$ & $0.12\pm0.11$\\
 \hline
 {\small DES-RG~708} &   {\small 0.60573} & 3$\rightarrow$2 & 215.351 & $<1.5$ & $<$2.6 & $<$0.23 & $<$0.09   &  $1.12^{+0.21}_{-0.17}$ & $0.13\pm0.11$ \\
 \hline 
 {\small COSMOS-FRI~16} &  {\small 0.9687} & 4$\rightarrow$3 & 234.185 & $<5.5$ & $<$18.8 & $<$7.2 & $<$1.5 &  $0.91^{+0.16}_{-0.14}$ & $0.32\pm0.22$ \\
 \hline 
 {\small COSMOS-FRI~31} &  {\small 0.9123} & 4$\rightarrow$3  & 241.036 & $<5.2$ & $<$15.8 & --- & $<$1.8   & $0.90^{+0.15}_{-0.13}$ & $0.33\pm0.24$ \\
 \hline
 {\small COSMOS-FRI~70} &  {\small 2.625} & 7$\rightarrow$6 &  222.525 & $0.69\pm0.31$ & $<6.6$ & $<$0.27 & $<$0.29 & $0.66^{+0.17}_{-0.13}$ & $0.71\pm0.28$ \\
 \hline
\end{tabular}
\caption{Molecular gas properties: (1) galaxy name;  (2) spectroscopic redshift as in Table~\ref{tab:radio_galaxies_properties0}; (3-4) CO(J$\rightarrow$J-1) transition and observer frame frequency; (5) CO(J$\rightarrow$J-1) velocity integrated flux; (6) molecular gas mass; (7) depletion time scale  $\tau_{\rm dep}=M({\rm H_2})/{\rm SFR_{24~\mu m}}$; (8) molecular gas to stellar mass ratio; (9-10) depletion time scale and molecular gas to stellar mass ratio for main sequence field galaxies \citep{Tacconi2018}. {Upper limits are reported at { 3$\sigma$}.}}
\label{tab:radio_galaxies_properties_mol_gas}
\end{adjustwidth}
\end{table*}

In the $\sim$(1.2-1.4)~mm wavelength range, the E230 receiver can offer 4$\times$4-GHz instantaneous bandwidth covered by the correlators. Of these four bands (UI, UO, LI, LO), we used only the lower side bands (LI, LO). 
The wobbler-switching mode was used for all the observations with a frequency of 0.5~Hz and a throw of either 60~arcsec or 120~arcsec, depending on the specific wind conditions found during the observations. The adopted wobbler throw is conservatively higher than the  size of our target sources, which is in fact less than a few arcseconds.

The Wideband Line Multiple Autocorrelator (WILMA) was used to cover the LI-4 GHz band in each linear polarization. The WILMA back-end gives a resolution of 2~MHz. We also simultaneously recorded the data with the Fast Fourier Transform Spectrometers (FTS), as a backup, at 200~kHz resolution, to cover the { 2$\times$4~GHz lower sidebands (LI and LO), for each linear polarization.}

The five target sources were observed for an on-source observing time of $\sim$71~hr in total, distributed among the five sources as follows: 6.8~hr (DES-RG~399), 17.8~hr (DES-RG~708), 14.0~hr (COSMOS-FRI~16), 12.4~hr (COSMOS-FRI~31), and 20.1~hr (COSMOS-FRI~70). 

Sources COSMOS-FRI~16 and 31 were observed during August 17-22, 2016, in bad weather conditions. Observations were carried out with an average precipitation water vapor (pwv) value of $\sim$12~mm, as well as high average system temperatures $T_{\rm sys}= 615$~K and 898~K, for COSMOS-FRI~16 and 31, respectively.
Sources DES-RG~399, DES-RG~708, and COSMOS-FRI~70 were observed during September 10-12, 2017, in very good weather conditions. Observations were carried out with average pwv values of 2.9~mm, 3.5~mm, and 1.5~mm, as well as average $T_{\rm sys}=219$~K, 327~K, and 294~K, for DES-RG~399, DES-RG~708, and COSMOS-FRI~70, respectively.

Only minor flagging was required: three scans, corresponding to 0.2~hr (on-source) of observations, were removed for COSMOS-FRI~31 because of bad weather conditions during their acquisition. The time reported above for COSMOS-FRI~31 corresponds to the net on-source time, where such scans have been removed. 

Data reduction and analysis were performed using the CLASS software of the GILDAS package\footnote{https://www.iram.fr/IRAMFR/GILDAS/}. The results are reported in Sect.~\ref{sec:IRAM30m_results}


\section{The wavelet-based Poisson Probability Method ($\mathit{w}$PPM)}\label{sec:wPPM}
We searched for megaparsec-scale overdensities around the radio galaxies in our sample using photometric redshifts of galaxies and the Poisson Probability Method \citep[PPM,][]{Castignani2014,Castignani2014b}, that we have improved using an approach based on the wavelet transform. We denote the  upgraded PPM as $\mathit{w}$PPM, where hereafter $\mathit{w}$ refers to the wavelet transform.
In  Sect.~\ref{sec:photoz_catalogs} we describe the photometric redshift catalogs used in this work. In Sect.~\ref{sec:PPM_proc} we outline the PPM procedure, while in Sect.~\ref{sec:wPPM_upgrade} we describe its wavelet-based upgrade. In Sect.~\ref{sec:Mpcscale_overdensities} we describe the $\mathit{w}$PPM results for the radio sources in our sample.

\subsection{Photometric redshift catalogs}\label{sec:photoz_catalogs}
We used photometric redshifts provided for the SDSS survey to search for megaparsec-scale overdensities around DES-RG~399 and 708. Photometric redshifts of SDSS~DR14 are estimated using a machine learning technique described in \citet{Csabai2007} and named kd-tree nearest-neighbor fit. We retrieved photometric redshifts within a rectangular field delimited by 40.2~deg.$<$R.A.$<$43.8~deg. and -1.8~deg.$<$Dec.$<$1.0~deg. Such a region corresponds approximately to the DES SN deep field n.~3 survey.  Following the SDSS~DR14 tutorial\footnote{http://www.sdss.org/dr14/algorithms/photo-z/} {we  considered sources with} \textsf{photoErrorClass}~= -1, 1, 2, or 3, which correspond to photometric redshifts with estimated rms errors in the range $\simeq(0.043-0.074)$. 

Concerning COSMOS-FRI~16, 31, and 70 we used the official COSMOS photometric redshift catalogs \citep{Ilbert2009,Laigle2016}. These catalogs are both obtained using the large multiwavelength photometric dataset provided for the COSMOS survey, which includes observations from the Hubble Space Telescope, Canada-France-Hawaii Telescope, and Subaru. The photometric redshift catalog by \citet{Laigle2016} includes also \textsf{YJHKs} photometry from the UltraVISTA survey \citep{McCracken2012}. Both photometric redshift catalogs were obtained by performing fits to the SEDs using LePhare \citep{Arnouts1999,Ilbert2006}. 

\subsection{The PPM procedure}\label{sec:PPM_proc}
We summarize here the basic steps of the PPM. We refer to \citet{Castignani2014b} for a detailed description.

\begin{figure*}[h!] \centering
\subfloat{\includegraphics[width=0.4\textwidth]{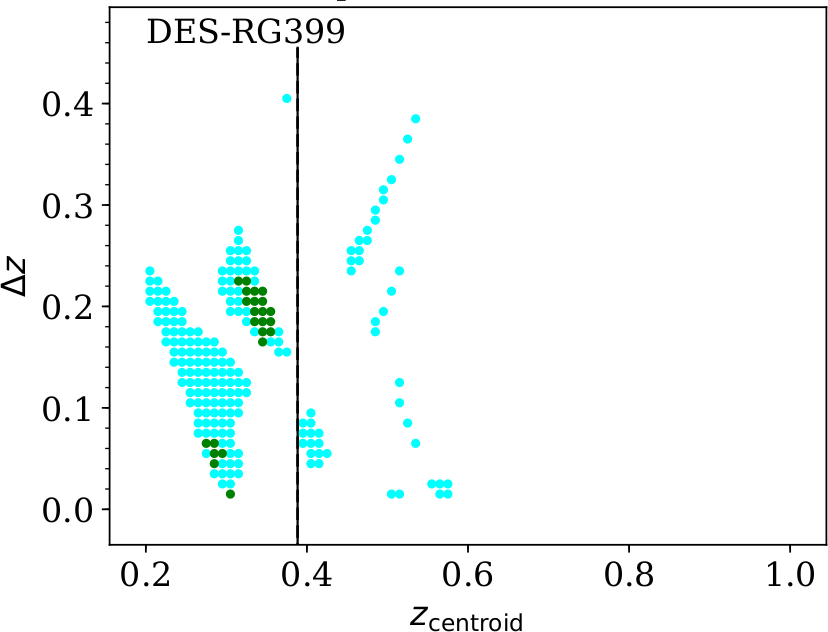}}
\subfloat{\hspace{0.1cm}\includegraphics[width=0.46\textwidth]{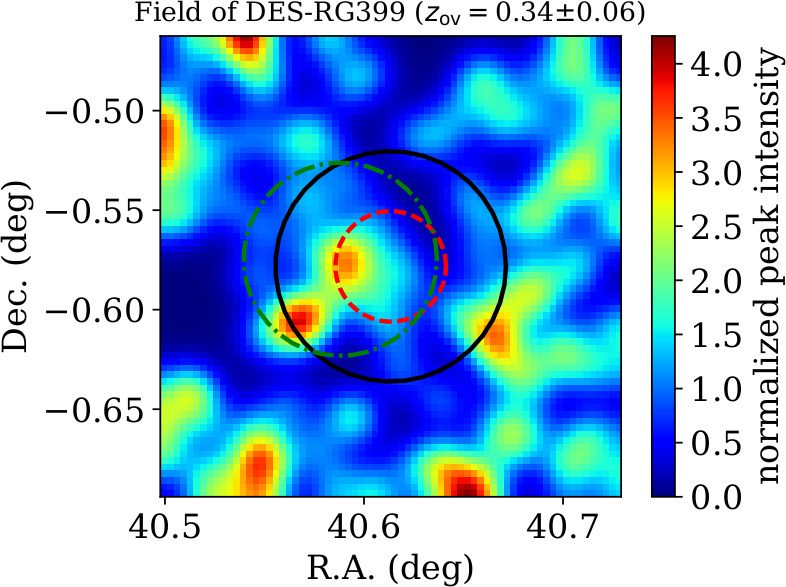}}\qquad
 \subfloat{\includegraphics[width=0.4\textwidth]{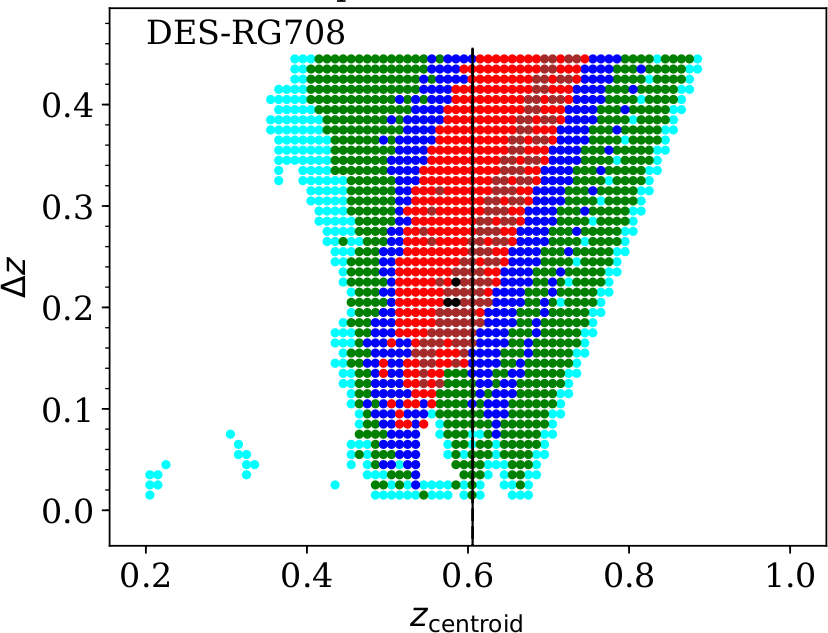}}
\subfloat{\hspace{0.1cm}\includegraphics[width=0.46\textwidth]{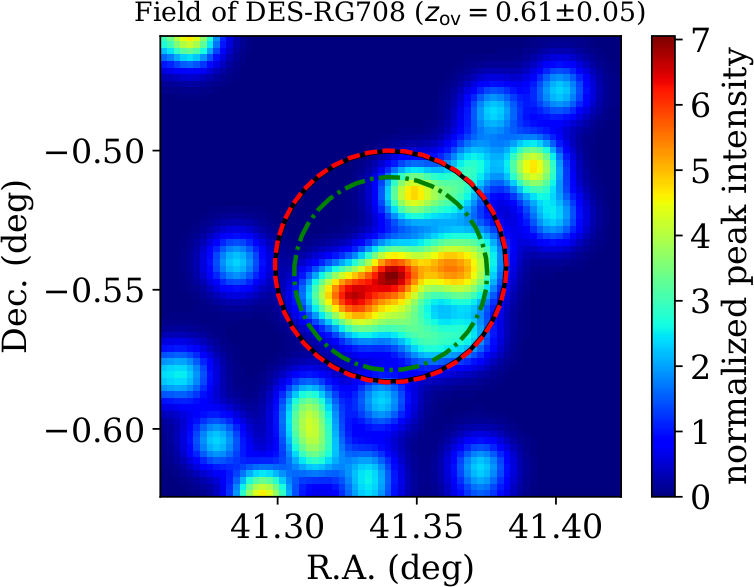}}\qquad
\caption{Left: PPM plots for DES-RG~399 and 708. In each plot the vertical solid line shows the spectroscopic redshift of each radio source. Colored dots refer to significance levels >2$\sigma$ (cyan), 3$\sigma$ (green), 4$\sigma$ (blue), 5$\sigma$ (red), 6$\sigma$ (brown), and 7$\sigma$ (black). Right: Gaussian density maps centered at the projected space coordinates of the radio galaxies. The pixel size is 1/16~Mpc while the Gaussian kernel has $\sigma=3/16$~Mpc. Sources with SDSS photometric redshifts between $z_{ov}-\overline{\Delta z}/2$ and $z_{ov}+\overline{\Delta z}/2$ were considered to produce the maps, where $z_{ov}$ and $\overline{\Delta z}$ are reported in Table~\ref{tab:cluster_properties}, for each overdensity. The solid black and dashed red circles are centered at the projected space coordinates of the radio source. The former has a (physical) radius of 1~Mpc, estimated at $z_{ov}$,  while the latter, with a radius $\mathcal{R}_{\rm PPM}$, denotes the region within which the PPM detects the overdensity. The dotted-dashed green circle is centered at the peak of the detection as found by the wavelet transform and has a radius $\mathcal{R}_{\mathit{w}}$.}
\label{fig:PPM_plots_DES}
\end{figure*}

\begin{figure*}[htpb] \centering
\subfloat{\includegraphics[width=0.4\textwidth]{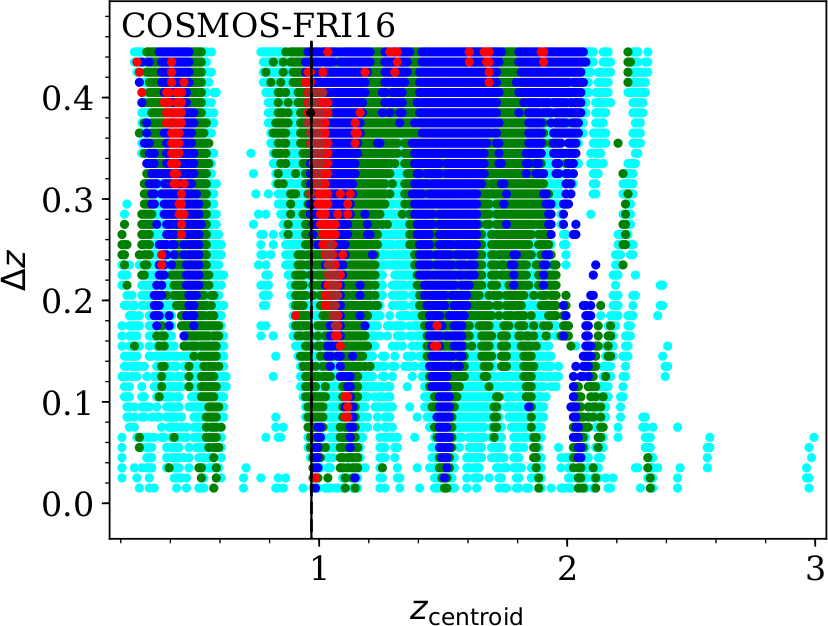}}
\subfloat{\hspace{0.1cm}\includegraphics[width=0.4\textwidth]{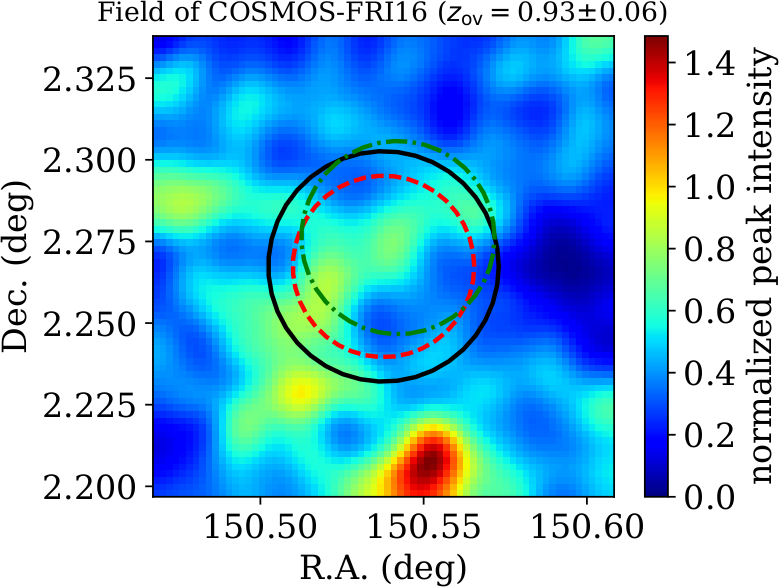}}\qquad
\subfloat{\hspace{7.5cm}\includegraphics[width=0.4\textwidth]{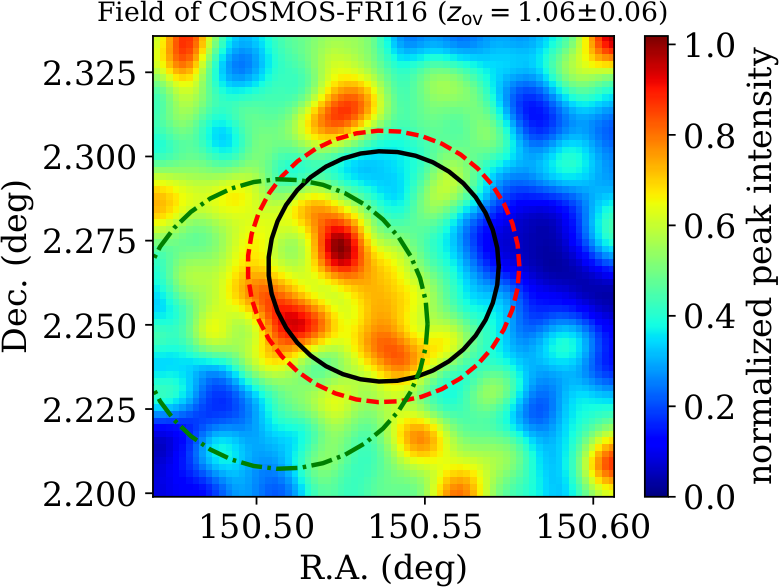}}\qquad
  \subfloat{\includegraphics[width=0.4\textwidth]{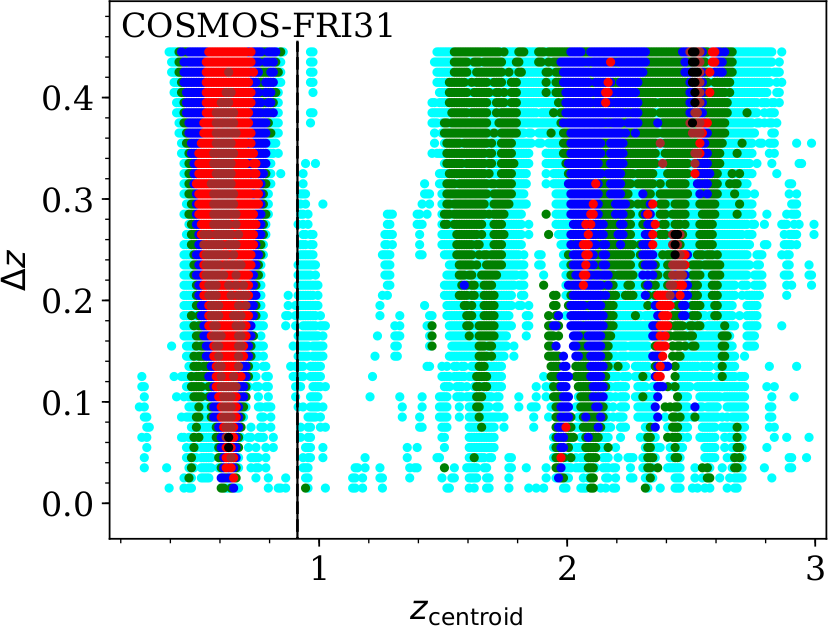}}
\subfloat{\hspace{0.1cm}\includegraphics[width=0.46\textwidth]{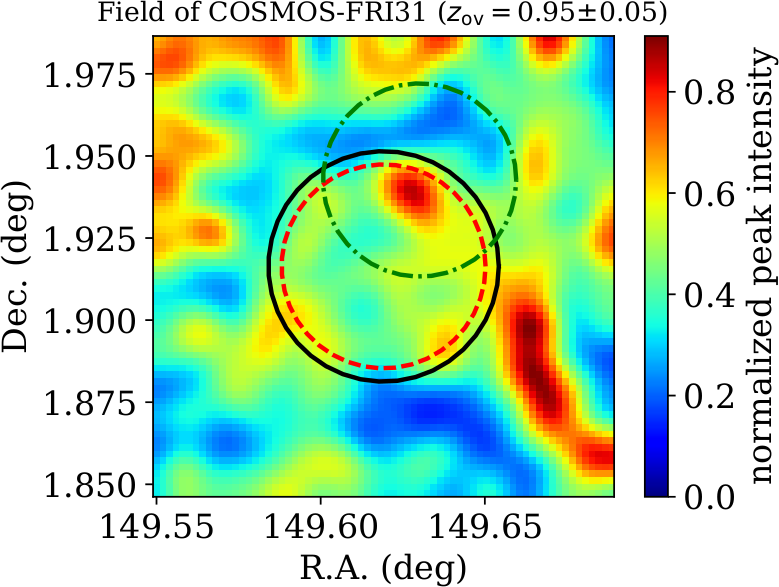}}\qquad
\subfloat{\includegraphics[width=0.4\textwidth]{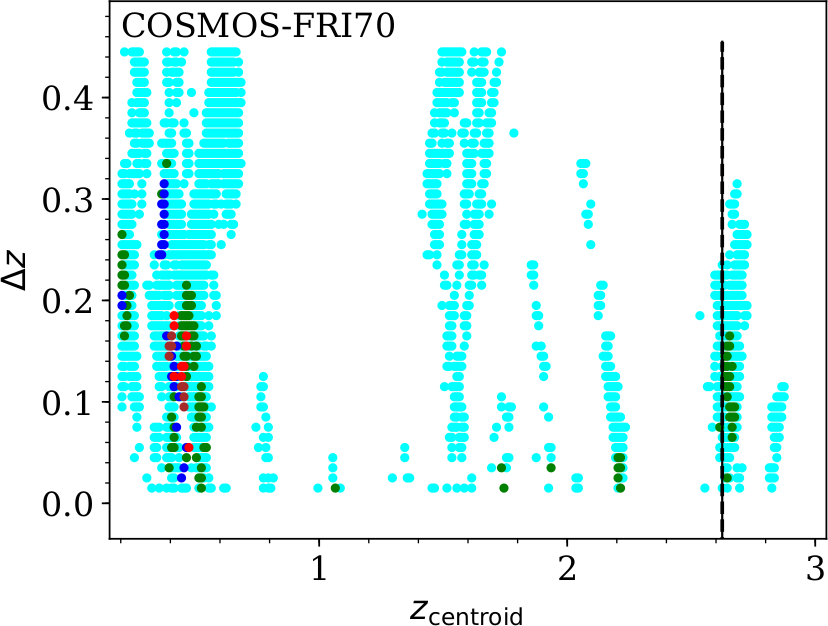}}
\subfloat{\hspace{0.1cm}\includegraphics[width=0.46\textwidth]{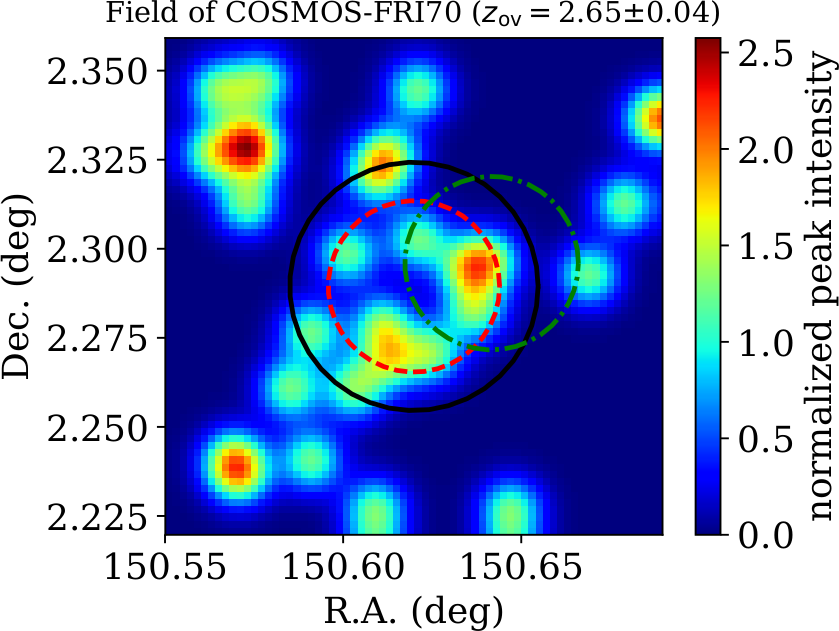}}\qquad
\caption{PPM plots (left) and wavelet density maps (right) associated with COSMOS-FRI~16, 31, and 70. The color code is analogous to that of Fig.~\ref{fig:PPM_plots_DES}. We have used photometric redshifts from \citet{Laigle2016} for the fields of COSMOS-FRI~16 and 31, and from \citet{Ilbert2009} for the field of COSMOS-FRI~70.}
\label{fig:PPM_plotsCOSMOS}
\end{figure*}

\begin{itemize}
\item We tessellate the projected space with a circle centered at the coordinates of the radio galaxy and a number of consecutive adjacent annuli. The annuli and the central circle have equal area {of 2.18~arcmin$^2$}. In particular the circle has a radius of 50~arcsec, which corresponds to physical scales in the range $\simeq$(0.3-0.4)~Mpc for the sources in our sample.\\
\item For each region {of the tessellation (the central circle and the consecutive annuli)}, we count the number of sources with photometric redshifts within an interval of length $\Delta z$ and  centered at the centroid redshift $z_{\rm centroid}$. The parameters $\Delta z$ and $z_{\rm centroid}$ uniformly span a grid of values that reflect the photometric redshift uncertainties and correspond to the redshift range of our interest, respectively.\\
\item  For each pair ($z_{\rm centroid};\Delta z$) we calculate the probability of the null hypothesis (i.e., no clustering) based on source number counts and Poisson statistics. To this aim the galaxy number density associated with each region of the tessellation is compared to that inferred from a sufficiently large control region. We chose rectangular control regions with subtended areas of 1.96~deg$^2$ and 1.44~deg$^2$ and with centers coincident to those of DES SN deep field n.~3 and COSMOS, respectively. The control regions are safely contained by the two surveys.
The procedure yields a number count excess significance associated with each pair ($z_{\rm centroid};\Delta z$), as well as a maximum radius within which the overdensity is detected. {Namely, the procedure selects the first consecutive regions starting from the central circle for which the probability of the null hypothesis is $\leq30\%$, for each pair ($z_{\rm centroid};\Delta z$)}.
\\
\item The PPM plots for the fields of the radio sources are derived. For each pair ($z_{\rm centroid};\Delta z$) the detection significance defined in the previous step is plotted.\\
\item As noted in \citet{Castignani2014,Castignani2014b} noisy features are present in the PPM plots because of photometric redshift uncertainties and shot noise associated with photometric redshift number counts. These noisy patterns might lead to spurious overdensity-to-radio-galaxy associations, especially in the case where the redshift of the radio galaxy is the photometric one.  However, at variance with the original PPM procedure, we decided not to use any Gaussian filter to eliminate high-frequency noisy patterns. This choice was motivated by the fact that we know the spectroscopic redshift of each radio galaxy. Such knowledge minimizes the chance of a spurious association between the radio source and the overdensities detected by the PPM along the line of sight. Furthermore, as also discussed in Sect.~\ref{sec:Mpcscale_overdensities}, our choice allows us to detect low-S/N overdensities that otherwise would be filtered-out by the smoothing procedure.\\
\item We have also improved the original PPM procedure by searching for an optimal redshift bin $\Delta z=\overline{\Delta z}$ which maximizes the overdensity significance, at a redshift $z_{\rm centroid}$  equal to the spectroscopic redshift of the radio galaxy. The search for a specific $\Delta z$ is motivated by the knowledge of the spectroscopic redshift of each radio source, with the aim to find an optimal overdensity significance and to minimize the risk of nondetection of the overdensity.
Inspection of the PPM plots in Figs.~\ref{fig:PPM_plots_DES} and \ref{fig:PPM_plotsCOSMOS} (left) yielded $\overline{\Delta z}\simeq0.15$, for COSMOS-FRI~70, and  $\overline{\Delta z}\simeq0.2$ for DES-RG~399, DES-RG~708, COSMOS-FRI~16, and COSMOS-FRI~31. The adopted values of $\overline{\Delta z}$ are therefore fairly independent of the specific redshift considered. As noted in \citet{Castignani2014,Castignani2014b} the overdensity patterns are in fact fairly stable along the y-axis, that is, with respect to different $\overline{\Delta z}$ values. Conservatively assuming a fiducial statistical photometric redshift uncertainty $\sigma(z)=\sigma_0(1+z)$, with $\sigma_0\simeq0.03$ for both SDSS and COSMOS surveys, we note that the adopted $\overline{\Delta z}$ values safely correspond to $\overline{\Delta z}/\sigma(z)$ ratios of the order of unity and equal to 4.8,  4.1,  3.4,  3.5, and 1.4, for DES-RG~399, DES-RG~708, COSMOS-FRI~16, COSMOS-FRI~31, and COSMOS-FRI~70, respectively.\\

\item At fixed $\overline{\Delta z}$, as found in the previous step, we have applied a peak-finding algorithm to the PPM plot. Such a procedure belongs to a more general context known as Morse theory and has been developed for our discrete case. Overdensities are found if they are associated with an interval  at least ${\delta z _{\rm centroid} = 0.03}$ in length on the redshift axis $z_{\rm centroid}$, at a given significance threshold $>2\sigma$.  Overdensities separated by less than 0.01 along the redshift axis $z_{\rm centroid}$ are also merged. 
By iteratively increasing the significance threshold the procedure provides us (1) the overdensity detection significance, (2) an estimate for the redshift ($z_{ov}$) of the overdensity, (3) an estimate for the (proto-)cluster core size ($\mathcal{R}_{\rm PPM}$), and (4) a rough estimate for the overdensity richness ($N_{\rm selected}$).\\
\item For each PPM plot any overdensity that is located at a redshift consistent with that of the corresponding radio source itself is associate with the radio galaxy, following the prescriptions described in \citet{Castignani2014}. Multiple overdensity associations are not excluded.\\
\end{itemize}


\subsection{The wavelet-based upgrade of the PPM}\label{sec:wPPM_upgrade}
The PPM does not search for overdensities blindly, but relies on a positional prior, that  is, the projected space coordinates of the radio galaxy. In particular the PPM exploits this positional prior to partially overcome the limitation due to low-number count statistics and shot noise, that usually affect distant cluster searches based on galaxy number counts.  As stressed in \citet{Castignani2014,Castignani2014b} such an exploitation is performed by privileging an accurate photometric redshift sampling to the detriment of a less sophisticated projected space tessellation. 

In this work we aim at improving the ability of the method to locate and characterize the overdensity in the projected space, once its redshift is determined by the PPM. As further described in the following this improvement is made by applying a 2D wavelet transform. 


\begin{itemize}
 \item For each overdensity, detected at a redshift $z_{ov}$, sources with photometric redshifts within $z_{ov}-\overline{\Delta z}/2$ and $z_{ov}+\overline{\Delta z}/2$ are considered. We used such sources to produce a 2D map with a pixel size of 1/16~Mpc. We then filtered the 2D map with the task \textsf{mr\_filter} within the multiresolution package MR/1 \citep{Starck1998}. The \textsf{mr\_filter} is based on the wavelet transform. We ran it to detect structures including a treatment of the Poisson noise and an iterative multiresolution thresholding down to 2$\sigma$, which is consistent with the value adopted by the PPM to define overdensities.\footnote{The adopted \textsf{mr\_filter} command is:
 ``\textsf{mr\_filter -m 10 -i 3 -s 10.,3.,2.,2. -n 5 -f 3 -K -C 2 -p -e 0 -A input output}'', where \textsf{input} and \textsf{output} denote the input and output .fits files corresponding to the original and wavelet-transformed maps, respectively.} Such a wavelet-based procedure is similar to that used by the WaZP cluster finder \citep{Benoist2014,Dietrich2014}.\\
 \item  From each wavelet map we then selected the highest peak falling within a circle centered at the projected coordinates of the radio galaxy and with a radius equal to $\mathcal{R}_{\rm PPM}$. We define as $\theta_{ov}$ the projected separation (i.e., the miscentering) of the radio galaxy coordinates with respect to the overdensity peak, as found by \textsf{mr\_filter}. This procedure aims at locating, in the projected space, the peak of the overdensity associated with the radio galaxy. For this reason, we limited the search to a circle of $\mathcal{R}_{\rm PPM}$ radius, which defines the overdensity found by the PPM. Extending the search to a larger region would possibly include peaks that are not physically associated with the radio galaxy. Distant (proto-)clusters can in fact exhibit a complex structure. For example, still-forming proto-clusters may extend up to $\sim$10~Mpc \citep[see e.g.,][for a review]{Overzier2016}. \\
\item A refinement of the estimate of the overdensity size is then derived as the minimum projected distance from the wavelet peak, selected as in the previous step, at which the wavelet map is reduced to one hundredth of the peak value. We denote such a size as $\mathcal{R}_{\mathit{w}}$. \\
\end{itemize}



\section{Results}\label{sec:results}
\subsection{Stellar masses}
{For all five sources in our sample we estimated the stellar masses from the SEDs of Sect.~\ref{sec:SEDmodeling} using a standard LePhare procedure, similarly to previous work \citep[e.g.,][]{Laigle2016}.}

Stellar masses were estimated by fitting the photometric data points with synthetic templates  of three elliptical galaxies from the PEGASE2 library \citep{Fioc_RoccaVolmerange1997}. A \citet{Rana_Basu1992} initial mass function (IMF) was assumed.

Concerning COSMOS-FRI~16, 31, and 70, B13 report stellar masses of $\log(M/M_\star)=10.74$, 10.75, and 10.65. They assumed several IMFs in their modeling \citep{Salpeter1955,Kroupa2001,Chabrier2003}.
From the results reported in Table~\ref{tab:radio_galaxies_SEDproperties} we observe that these mass estimates are fairly consistent with those estimated independently in this work. In fact stellar-mass estimates rely on stellar-population synthesis models and have statistical uncertainties of $\sim(0.10-0.14)$~dex \citep[e.g.,][]{Roediger_Courteau2015}. An additional uncertainty of $\sim$0.25~dex may be added because of the unknown IMF \citep{Wright2017}, yielding a typical uncertainty of $\sim0.3$~dex, which is adopted for all stellar masses estimated in this work and reported in Table~\ref{tab:radio_galaxies_SEDproperties}.

\subsection{Star formation rates}\label{sec:SFR}
By integrating the SED best-fit model for the dust between 8  and 1000$\mu$m in the rest frame we derived an estimate for the FIR luminosity that we converted into an estimate for the star formation rate (SFR) using the \citet{Kennicutt1998} relation. We denote such SFR estimates as SFR$_{\rm SED}$, where the subscript SED stands for the fact that the SFR is estimated using the FIR SED best-fit model.

Our SFR$_{\rm SED}$ estimates are indeed upper limits, since in the FIR $>24~\mu$m observer frame domain we only have upper limits to the fluxes. Therefore we also estimated the SFRs using the available $\sim$24~$\mu$m (Spitzer MIPS or WISE W4) fluxes. We denote such SFR estimates as SFR$_{\rm 24\mu m}$, where the subscript \emph{24~$\mu$m} refers to the fact that the SFR is estimated using the $\sim$24~$\mu$m fluxes. To estimate SFR$_{\rm 24\mu m}$ we adopted the same procedure described in \citet{Castignani2018} that is adapted from that of \citet{McDonald2016}, as outlined below.


{We estimated the $\lambda=24~\mu$m rest-frame luminosities $\lambda L_\lambda$ from the observer-frame $\sim$24-$\mu$m fluxes by assuming a power-law model, $L_\lambda\propto\lambda^\gamma$, with $\gamma=2.0\pm0.5$ \citep{Casey2012}. We have incorporated the uncertainties in both $\gamma$ and the observed 24-$\mu$m fluxes by using $M=100,000$ values drawn from Gaussian distributions centered at the mean values and with standard deviations equal to the associated uncertainties. Subsequently, we adopted the \citet{Calzetti2007} relation to convert the  24-$\mu$m rest-frame luminosities into the SFR$_{\rm 24\mu m}$ estimates. The SFR$_{\rm 24\mu m}$ values and uncertainties were finally estimated as the  medians and the 68.27\% confidence levels derived from the $M$ realizations, respectively.}

For COSMOS-FRI~31, we only have upper limits in the FIR. We therefore derived a 3$\sigma$ upper limit to the SFR$_{\rm 24\mu m}$.   Hereafter we adopt SFR$_{\rm 24\mu m}$ as our fiducial estimate for the SFR. The SFR$_{\rm 24\mu m}$ is in fact always lower than and therefore consistent with the SFR$_{\rm SED}$ upper limits. 

{Estimating the SFR from the observer frame 24-$\mu$m flux could be problematic for distant sources, because in the rest frame such a flux corresponds to the emission at shorter wavelengths, namely 24~$\mu$m$/(1+z)$, where $z$ is the source redshift. Nevertheless, we stress that the observer frame 24-$\mu$m flux is adopted in the literature as a proxy for the SFR, in the absence of other SFR estimators, also for distant galaxies, such as in the recent work by \citet{Wang2018} on cluster galaxies at  $z\sim2.5$  with detections in CO(1$\rightarrow$0).}

The specific SFR estimates for our sources are in overall agreement, within the reported uncertainties, with the {empirical values} by \citet{Speagle2014} for main sequence (MS) field galaxies with similar redshift and stellar mass, as reported in Table~\ref{tab:radio_galaxies_SEDproperties}. We point out that our SFR estimates are also supported by a hint for a UV bump, possibly associated with star formation, observed by B13, in the SED of COSMOS-FRI~16, 31 and 70 (Fig.~\ref{fig:SEDs}).

We stress that our radio sources, in particular DES-RG~399 and 708, have nearby companions clearly visible in their optical images (see e.g., Fig.~\ref{fig:DES_RG_images}), located at an angular separation consistent with the WISE and Spitzer MIPS positional accuracy of $\sim6$~arcsec. It is therefore possible that such companions might contaminate the observed IR emission, resulting in biased-high SFR estimates. 
Similarly, active galactic nucleus (AGN) torus and/or synchrotron emission could contaminate the observed FIR spectrum, even if at such wavelengths the AGN contribution is typically limited to $\lesssim20\%$ for distant star forming galaxies \citep{Donley2012,Pozzi2012,Delvecchio2014}. We also note that our photometric data set does not allow us to properly break the degeneracy between the synchrotron, torus, and dust emission. 

\begin{figure*}[h!]\centering
\subfloat{\includegraphics[width=0.32\textwidth]{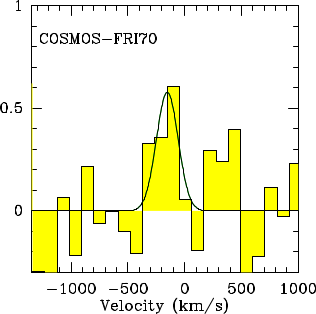}}
\subfloat{\hspace{0.1cm}\includegraphics[width=0.3\textwidth]{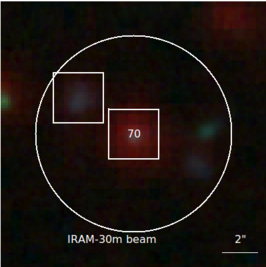}}
\caption{Left: Baseline-subtracted spectrum of COSMOS-FRI~70 obtained with the IRAM-30m. The solid curve shows the Gaussian fit to the CO(7$\rightarrow$6) emission line. In the y-axis we show $T_{\rm mb}$ in units of millikelvin. Right: RGB image centered at the coordinates of COSMOS-FRI~70 and obtained using Spitzer 3.6$\mu$m \citep{Sanders2007}, Subaru \textsf{r}-, and Subaru
\textsf{B}-band images \citep{Taniguchi2007} for the R, G, and B channels, respectively. The two squares denote the locations of COSMOS-FRI~70 and its northern companion with $z_{\rm phot}=2.07^{+0.09}_{-0.41}$ from the \citet{Laigle2016} catalog. The white circle shows the IRAM-30m beam. North is up.}
\label{fig:IRAM30m_results}
\end{figure*}

\subsection{Molecular gas properties}\label{sec:IRAM30m_results}
We describe in this section the results obtained with our IRAM-30m observations.
All our targets were unresolved by our observations, with a beam of $\sim$11~arcsec~$\big(\frac{210~{\rm GHz}}{\nu_{\rm obs}}\big)$ at observer frame frequency $\nu_{\rm obs}$ \citep{Kramer2013}.

{A hint of a tentative CO(7$\rightarrow$6) detection is found at $2.2\sigma$} for COSMOS-FRI~70. 
In Fig.~\ref{fig:IRAM30m_results} we report the spectrum obtained for COSMOS-FRI~70 as well as an RGB image of the radio source and the IRAM-30m field of view. Visual inspection of Fig.~\ref{fig:IRAM30m_results} (right) reveals several sources within the IRAM-30m beam. Among them the galaxy with the smallest angular separation (i.e., 3.7~arcsec) from COSMOS-FRI~70 has a photometric redshift from the  COSMOS2015 catalog \citep{Laigle2016} equal to $z_{\rm phot}=2.07^{+0.09}_{-0.41}$ , which is (marginally) consistent with the spectroscopic redshift of COSMOS-FRI~70, considering the large photometric redshift uncertainties. Here the subscript \emph{phot} stands for photometric.
We cannot exclude the possibility that {the tentative CO(7$\rightarrow$6)
emission observed in the spectrum} is due to our target COSMOS-FRI~70 and also possibly  to its nearby companion, both unresolved by the IRAM-30m. {We stress that confirmation with higher-S/N data is needed, to claim a detection.}

For the other four sources in the sample, DES-RG~399, DES-RG~708, COSMOS-FRI~16, and COSMOS-FRI~31, we reached rms noise levels for the antenna temperature (Ta$^\ast$) equal to 0.2, 0.2, 0.8, and 0.7~mK, respectively, within the entire 4-GHz bandwidth and at 300~km/s resolution.  We used the rms noise levels to set { 3$\sigma$} upper limits.

In Table~\ref{tab:radio_galaxies_properties_mol_gas} we report the results of our analysis, where standard efficiency corrections have been applied  to convert i) Ta$^\ast$ into the main beam temperature $T_{\rm mb}$ and then ii) $T_{\rm mb}$ into the corresponding CO line flux, where a 5-Jy/K conversion is used.

{In Table 4 we report $3\sigma$ upper limits to the total molecular gas mass for all five radio sources in our sample, including COSMOS-FRI~70, as described in the following.}

\begin{table*}[htb]
\begin{center}
\begin{tabular}{ccccccc}
\hline\hline
 Galaxy ID &  $\log(M_\star/M_\odot)$ & $L_{\rm FIR}$ & SFR$_{\rm SED}$  & SFR$_{\rm 24\mu m}$ &  sSFR  & sSFR$_{\rm MS}$ \\
   &  & ($L_\odot$) &  ($M_\odot/{\rm yr}$) &  ($M_\odot/{\rm yr}$) & (Gyr$^{-1}$) & (Gyr$^{-1}$) \\ 
 (1) & (2) & (3) & (4) & (5) & (6) & (7)  \\
 \hline
 {\small DES-RG~399} & 10.96 & $<$3.9e+11 & $<$68 & $21^{+10}_{-9}$ & $0.23^{+0.48}_{-0.16}$  &  0.11 \\ 
 \hline
 {\small DES-RG~708} & 11.45 & $<$1.51e+12 & $<$263 & $112^{+68}_{-58}$ & $0.40^{+0.83}_{-0.30}$ &  0.13 \\
 \hline 
 {\small COSMOS-FRI~16} &  11.09 & $<$2.7e+11 & $<$48 & $26^{+9}_{-7}$ & $0.21^{+0.44}_{-0.14}$ & 0.40 \\  
 \hline
 {\small COSMOS-FRI~31} &  10.94 & $<$4.1e+11 & $<$72 & $<42$ & $<0.48$ & 0.38 \\
 \hline
 {\small COSMOS-FRI~70} & 11.36 & $<$1.2e+13 & $<$2050 & $245^{+205}_{-112}$ &  $1.07^{+2.32}_{-0.78}$ & 1.30 \\
 \hline
 \end{tabular}
\end{center}
 \caption{Results of the SED modeling: (1) galaxy name; (2) stellar mass; (3) FIR (8-1000)$\mu$m luminosity; (4) SFR estimated integrating the best fit SED model between (8-1000)$\mu$m in the rest frame; (5) SFR estimated from the available $\sim$24~$\mu$m (WISE W4 or Spitzer MIPS) flux and the \citet{Calzetti2007} relation, for COSMOS-FRI~31 the 2$\sigma$
upper limit is reported; (6) specific SFR derived as sSFR=SFR$_{\rm 24\mu m}/M_\star$, where the uncertainties are obtained 
by propagating those of of SFR$_{\rm 24\mu m}$ and those $\sim$0.3~dex of $M_\star$; (7) specific SFR from \citet{Speagle2014} for main sequence (MS) field galaxies of redshift and stellar mass equal to those of our targets.}
\label{tab:radio_galaxies_SEDproperties}
\end{table*}

\begin{table*}[htb]
\begin{center}
\begin{tabular}{cc|l}
\hline\hline
(1) & Galaxy ID & COSMOS-FRI~70\\
\hline
(2) & $z_{\rm CO(7\rightarrow6)}$ & $2.623\pm0.001$ \\
\hline
(3) &    FWHM & $(225\pm123)$~km/s\\
\hline
(4) &    $L^{\prime}_{CO(7\rightarrow6)}$  & $(4.5\pm2.0)\times10^9~{\rm K~km~s}^{-1}~{\rm pc}^2$ \\
\hline
\end{tabular}
\end{center}
\caption{Additional IRAM-30m results for COSMOS-FRI~70: (1) galaxy ID; (2) spectroscopic redshift, (3) FWHM, and (4) velocity integrated luminosity inferred from the CO(7$\rightarrow$6) line.}
\label{tab:COSMOS_FRI70results}
\end{table*}

\subsubsection{Molecular gas mass}
We derived the CO(J$\rightarrow$J-1) luminosity $L^{\prime}_{\rm CO(J\rightarrow J-1)}$ from the velocity integrated CO(J$\rightarrow$J-1) flux $S_{\rm CO(J\rightarrow J-1)}\,\Delta\varv\ $ by using Eq.~(3) of \citet{Solomon_VandenBout2005}:
\begin{equation}
\label{eq:LpCO}
 L^{\prime}_{\rm CO(J\rightarrow J-1)}=3.25\times10^7\,S_{\rm CO(J\rightarrow J-1)}\,\Delta\varv\,\nu_{\rm obs}^{-2}\,D_L^2\,(1+z)^{-3}\,,
\end{equation}
where $\nu_{\rm obs}$ is the observer frequency of the CO(J$\rightarrow$J-1) transition, $D_L$ is the luminosity distance, and $z$ the redshift of the radio source.

By assuming a Galactic CO-to-H$_2$ conversion factor $X_{CO}\simeq2\times10^{20}$~cm$^{-1}$/(K km/s), that is,  $\alpha_{CO}=4.36~M_\odot\,({\rm K~km~s}^{-1}~{\rm pc}^2)^{-1}$, typical of MS galaxies \citep{Solomon1997,Bolatto2013}, {we estimated the $3\sigma$ upper limits} to the total molecular gas masses $M({\rm H_2})=\alpha_{CO}L^{\prime}_{CO(1\rightarrow0)}=\alpha_{CO}L^{\prime}_{CO(J\rightarrow J-1)}/r_{J1}$ {for the five radio galaxies in our sample.}
Here $r_{J1}= L^{\prime}_{CO(J\rightarrow J-1)}/L^{\prime}_{CO(1\rightarrow0)}$ is the excitation ratio. Concerning the CO(3$\rightarrow$2), CO(4$\rightarrow$3) and CO(7$\rightarrow$6) lines considered in this work we have assumed the following fiducial excitation ratios, namely, $r_{31}=0.55$ \citep{Devereux1994,Daddi2015}, $r_{41}=0.40$ \citep{Papadopoulos2000} , $r_{71}=0.39$ \citep{GonzalezLopez2017}, that are adopted in the literature for star forming galaxies.

\subsubsection{Star formation, main sequence, and depletion time}
{Among the five radio sources in our sample COSMOS-FRI~70 is the only one for which a hint of detection in CO is reported, at $2.2\sigma$. Interestingly, if we assume $M({\rm H}_2)=(5.0\times2.2)\times10^{10}~M_\odot$ inferred by our observations and corresponding to the formal $3\sigma$ upper limit of $M({\rm H}_2)<6.6\times10^{10}~M_\odot$  reported in Table~\ref{tab:radio_galaxies_properties_mol_gas},} the ${\rm SFR(24\mu{\rm m})}=(245^{+205}_{-112})~M_\odot$/yr estimated for COSMOS-FRI~70 (see Table~\ref{tab:radio_galaxies_SEDproperties}) agrees, within the errorbars, with the SFR expected from standard $M({\rm H_2})$ versus SFR relations. We used the relation between $M({\rm H_2})$ and the IR luminosity for both MS and color-selected star forming galaxies at $z\sim0.5-2.3$ found by \citet{Daddi2010} as well as that of \citet{Kennicutt1998} between the total FIR luminosity and the SFR.
This procedure yields an SFR$=(136\pm104)$~$M_\odot/$yr, where the reported uncertainties take into account those estimated for $M({\rm H_2})$ that are summed in quadrature to fiducial $\sim58\%$, that is, 0.25~dex, uncertainties associated with the calibration and the scatter of the adopted scaling relations \citep{Daddi2010}. 

{We used the SFRs and molecular-mass estimates to set $3\sigma$ upper limits to the depletion time scale, associated with the consumption of the molecular gas of the the radio sources in our sample, apart for COSMOS-FRI~31;}  this latter has upper limits to both $M({\rm H_2})$ and ${\rm SFR_{24~\mu m}}$ which did not allow us to estimate $\tau_{\rm dep}$ or its upper limit. Similarly, we set $3\sigma$ upper limits to the $M({\rm H}_2)/M_\star$ ratio for our radio sources.

For comparison we also estimated the depletion time $\tau_{\rm dep, MS}$ and the molecular gas to stellar mass ratio $\big(\frac{M({\rm H_2})}{M_\star}\big)_{\rm MS}$ for MS field galaxies with redshift and stellar mass equal to those of our target galaxies, {as found using the empirical relations provided by} \citet{Tacconi2018} . Our results are summarized in Table~\ref{tab:radio_galaxies_properties_mol_gas}, while in Table~\ref{tab:COSMOS_FRI70results} we report additional properties of COSMOS-FRI~70 obtained with our IRAM-30m observations.

\subsubsection{Continuum emission}
We also attempted to set an upper limit on the continuum emission of COSMOS-FRI~70 by using the available total 8-GHz bandwidth associated with the LI and LO channels of WILMA and FTS, respectively, for each polarization.  
COSMOS-FRI~70 is in fact the only source in our sample that is observed at a relatively short wavelength, in the rest frame, $\sim$372~$\mu$m, which is close to the $S_\nu$ dust emission peak at $\sim$100~$\mu$m in the rest frame. It is also the source for which we achieved the smallest rms noise with our IRAM-30m observations.
We obtained an rms of 0.1~mJy over the total 8-GHz bandwidth. The rms is about one order of magnitude lower than the absolute value of the observed baseline level, which is due to the atmospheric turbulence. Therefore we conclude that the faintness of our targets, combined with the significant intrinsic atmospheric instability at mm wavelengths, prevents us from setting robust upper limits to the continuum emission of our targets.

\subsection{Molecular gas properties of distant (proto-)cluster galaxies}
We compare the results found for our galaxies in terms of molecular gas content, SFR, and depletion time with those found by 
previous work on distant (proto-)cluster galaxies at redshifts of $z\sim0.2-5$. Similarly to previous work by \citet{Noble2017} and \citet{Castignani2018} we considered (proto-)cluster galaxies from the literature with estimates of both stellar and total gas masses, as well as estimates or upper limits to the SFR. 

We considered stellar masses $M_\star>10^{10}~M_\odot$  and SFR$<6$~SFR$_{\rm MS}$, where ${\rm SFR}_{\rm MS}$ is the SFR estimated using the {empirical recipe by} { \citet{Speagle2014}} for MS field galaxies of stellar mass and redshift equal to those of the galaxy considered.
Including galaxies significantly above the MS might lead to biased-high molecular-gas-to-stellar-mass ratios. Such galaxies are also usually associated with lower values for $\alpha_{\rm CO}$; see for example \citet{Noble2017} and \citet{Castignani2018} for a discussion. { As further outlined below most of the galaxies considered for the comparison, including our target galaxies, have SFR$\lesssim3$~SFR$_{\rm MS}$, and are formally consistent with being on the MS. For the sake of completeness we did not discard the remaining sources with $3<$SFR/SFR$_{\rm MS}<6$, given also the large uncertainties associated with the SFR.}

In the following we distinguish between galaxies at redshifts lower and higher than $z\simeq2$, which usually denotes in the literature the fiducial separation between galaxy clusters and proto-clusters, respectively. 

\subsubsection{Cluster galaxies at  $0.2\lesssim z\lesssim2.0$}
We considered CO detections of distant, that is, $z\sim0.2-2$, cluster galaxies at $z\sim0.2$ \citep{Cybulski2016}; $z\sim0.4-0.5$ \citep{Geach2011,Jablonka2013}; $z\sim1.1-1.2$ \citep{Wagg2012,Castignani2018}; $z\sim1.5-1.7$ \citep{Aravena2012,Rudnick2017,Webb2017,Noble2017,Noble2018,Hayashi2018}; and $z\simeq2.0$ \citep{Coogan2018}. We also included the serendipitous CO detections reported by \citet{Kneissl2018} and associated with galaxies likely belonging to the cluster candidate PLCK~C073.4-57.5 at $z=1.54$.

{Requiring SFR$<6$~SFR$_{\rm MS}$ and $M_\star>10^{10}~M_\odot$ yields 52 sources} over 16 clusters. 
In the following we consider proto-cluster galaxies at $z>2$.

\subsubsection{Cluster galaxies at  $2\lesssim z\lesssim5$}
We searched the literature for CO detections of cluster galaxies at $z>2$. To this aim we considered the list of detections reported in \citet{Dannerbauer2017}. Namely, we include the proto-cluster galaxies with CO detections at $z=2.148$ \citep{Dannerbauer2017}; $z=2.41$ \citep{Ivison2013}; $z=2.51$ \citep{Tadaki2014}; $z=4.05$ \citep[GN20 proto-cluster,][]{Tan2014}; $z=5.2$ \citep[galaxy HDF850.1,][]{Walter2012}; and $z=5.3$ \citep[AzTEC-3 proto-cluster,][]{Riechers2010}. For the cluster galaxy HDF850.1 at $z=5.2$ we assumed the stellar-mass estimate $M_\star=(1.3\pm0.6)\times10^{11}~M_\odot$ by \citet{Serjeant_Marchetti2014}.

Furthermore we also considered the Spider Web Galaxy, namely MRC~1138-262. It is a radio galaxy at $z=2.161$ hosted in a proto-cluster, with CO(1$\rightarrow$0) detection reported by \citet{Emonts2013,Emonts2016} and  $M_\star\simeq10^{12}~M_\odot$ \citep{Hatch2009}, for which we assumed a fiducial uncertainty of $\sim0.5$~dex.

{We considered also the following recent CO observations by \citet{Lee2017}, \citet{Ginolfi2017}, and \citet{Wang2018}.}
\citet{Lee2017} detected in CO(3$\rightarrow$2) a sample of seven galaxies associated with a proto-cluster at $z=2.49$ around the powerful high-$z$ radio galaxy 4C~23.56. 
{\citet{Ginolfi2017} reported the CO(4$\rightarrow$3) emission from Candels-5001, a massive galaxy ($M_\star\sim1.9\times10^{10}~M_\odot$) at $z=3.47$ hosted in a proto-cluster candidate.
\citet{Wang2018} found CO(1$\rightarrow$0) in a sample of 14 galaxies belonging to CLJ1001, the most distant X-ray  cluster, at $z=2.51$. The authors do not report SFR estimates for these 14 sources. Therefore, consistently with the procedure adopted in this work, we converted the total IR luminosities reported by \citet{Wang2018} for the 14 sources into SFR estimates using the \citet{Kennicutt1998} relation.}

This procedure yielded {26 sources} with SFR$<6$~SFR$_{\rm MS}$ and $M_\star>10^{10}~M_\odot$, over {seven (proto-)clusters} at $z>2$. We note that \citet{Ivison2013} reported CO(1$\rightarrow$0) detections associated with four hyper-luminous IR  galaxies belonging to the proto-cluster HATLAS~J084933 at $z=2.41$  and with an estimated SFR in the range $\sim(640-3400)~M_\odot$/yr. We did not select such sources since they do not satisfy the SFR$<6$~SFR$_{\rm MS}$ criterion. For the same reason we also excluded the sub-millimeter proto-cluster galaxy AzTEC at $z=5.3$ and with an estimated SFR of $\sim1800~M_\star$/yr \citep{Riechers2010}.\\

By combining together both $z\lesssim2$ and $z\gtrsim2$ (proto-)cluster galaxies our selection yielded a total of {78 galaxies} over 23 (proto-)clusters with CO detections from the literature. To them we have also added our 5 target radio galaxies in dense megaparsec-scale environment (see Sect.~\ref{sec:Mpcscale_overdensities}), implying a total of {83} galaxies over 28 (proto-)clusters.

We stress here that the cluster galaxy GAL0926+1242-B that i) belongs to the $z=0.49$ cluster CL0926+1242, ii) has a stellar mass $M_\star=3.8\times10^9~M_\odot$, and iii) has been detected in CO by \citet{Jablonka2013} is the only source that was discarded because it does not satisfy the stellar mass criterion $M_\star>10^{10}~M_\odot$, despite having SFR$<6$~SFR$_{\rm MS}$.

\subsubsection{Additional distant (proto-)cluster galaxies with CO detections}
Additional $z>0.2$ (proto-)cluster galaxies have been detected in CO by previous studies, but we have not included them in our analysis because they lack stellar-mass estimates from the literature. For the sake of completeness we list such sources in the following.

\citet{Casasola2013} detected CO(2$\rightarrow$1) from an AGN belonging to a distant cluster around the powerful radio galaxy 7C~1756+6520 at $z\simeq1.4$. The CO-emitting AGN is located at a projected separation of $\sim$0.8~Mpc from the radio galaxy.
Interestingly,  \citet{Ivison2012} detected CO(1$\rightarrow$0) in two distant $z\sim3-3.5$ powerful radio galaxies (i.e., B3 J2330+3927 and 6C~1909+72) hosted in dense  megaparsec-scale environment \citep{Stevens2003}. 
\citet{Hayatsu2017} detected a CO(9$\rightarrow$8) emitter at $z=3.1$, which was identified by the authors as a member of a known proto-cluster in the SSA22 region \citep{Steidel1998}.
\citet{Oteo2018} and \citet{Miller2018} recently reported several CO detections in the core of two proto-clusters at $z= 4.002$ and $z= 4.3$, respectively.


\begin{figure*}[]\centering
\subfloat{\includegraphics[width=0.346\textwidth]{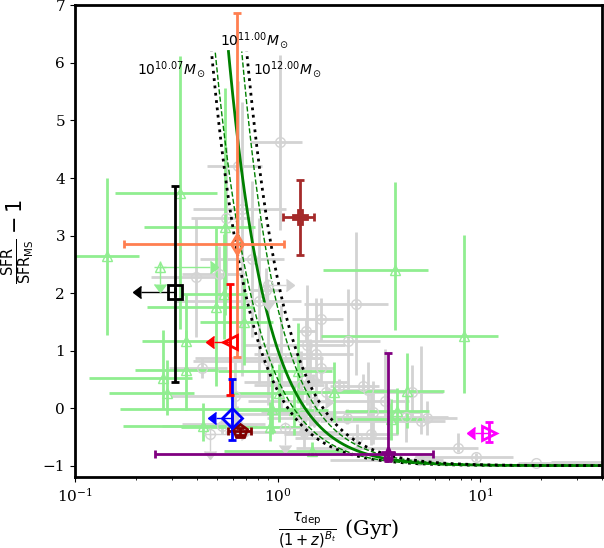}}
\subfloat{\hspace{0.2cm}\includegraphics[width=0.5\textwidth]{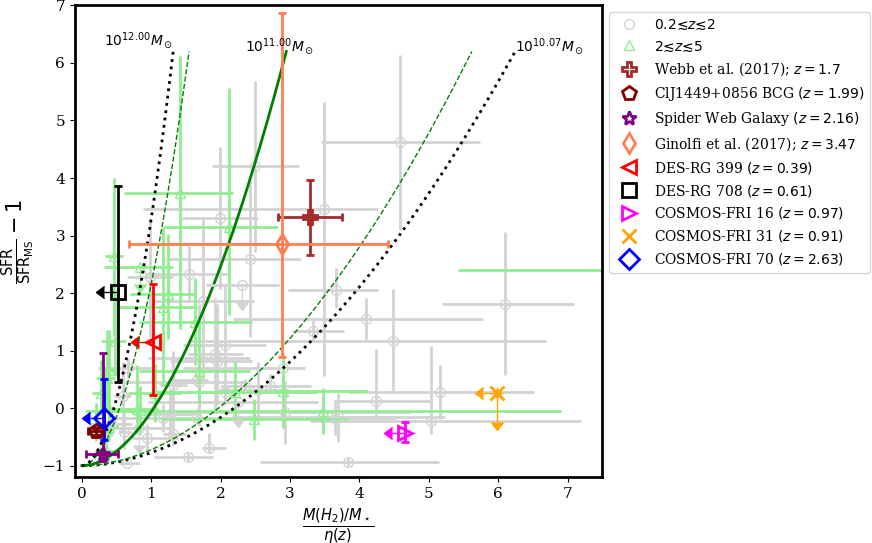}}\\
\caption{Fractional offset from the star-forming MS as a function of the molecular gas depletion timescale (left) and { molecular-gas-to-stellar-mass ratio} (right); (proto-)cluster galaxies at $0.2\lesssim z\lesssim5$ detected in CO are shown. 
In both panels the solid green curve shows the scaling relation for field galaxies found by \citet{Tacconi2018} for galaxies with $\log(M_\star/M_\odot)$=11 {and an effective radius equal to the mean value found by \citet{vanderWel2014}  for star forming galaxies for given $z$ and $M_\star$}. The {green} dashed lines show the statistical 1$\sigma$ uncertainties in the model. { The dotted black lines are the same scaling relation as the solid green lines, but for different stellar masses $\log(M/M_\star)=$~10.07 and 12, that represent the stellar mass range associated with the data points.}}
\label{fig:gas_properties}
\subfloat{\includegraphics[width=0.5\textwidth]{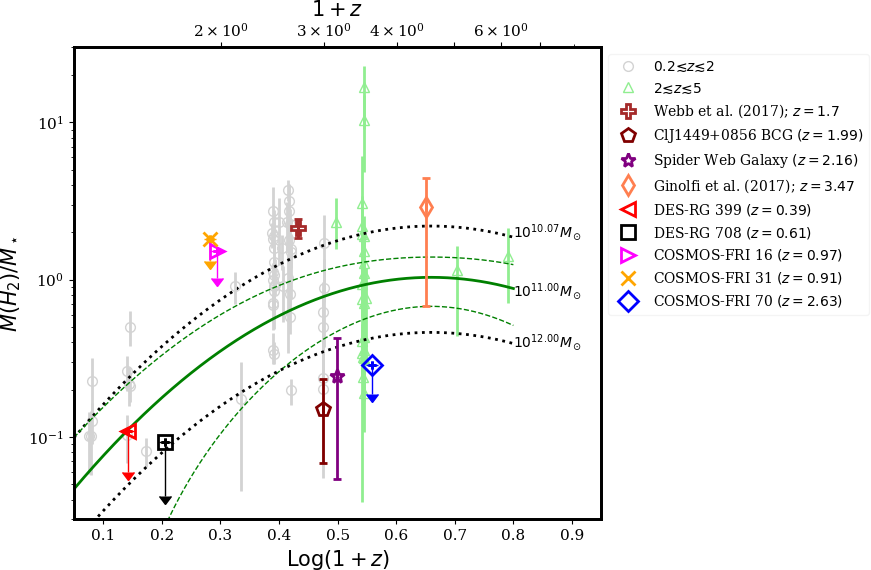}}\\
\caption{Evolution of the {molecular-gas-to-stellar-mass ratio} as a function of the redshift for (proto-)cluster galaxies at $0.2\lesssim z\lesssim5$ detected in CO. The color code is analogous to that of Fig.~\ref{fig:gas_properties}. 
The solid{ green} curve is the scaling relation found by \citet{Tacconi2018} for field galaxies in the MS and  with $\log(M_\star/M_\odot)$=11, { where an effective radius equal to the mean value found by \citet{vanderWel2014} for star forming galaxies for given $z$ and $M_\star$ is assumed.} The {green} dashed lines show the statistical 1$\sigma$ uncertainties in the model. { The dotted black lines show the same scaling relation as the solid green line, but for different stellar masses $\log(M/M_\star)=$~10.07 and 12, that correspond to the stellar-mass range spanned by the data points.} }
\label{fig:gas_fraction}
\end{figure*}

\subsubsection{Molecular gas diagrams}\label{sec:molecular_gas_diagrams}
In Figs.~\ref{fig:gas_properties} and \ref{fig:gas_fraction} we report observational results concerning the aforementioned sample of 83 cluster galaxies from both the literature (78 sources) and this work (5 sources). 

In Fig.~\ref{fig:gas_properties} we show the fractional offset from the star forming MS as a function of the molecular gas depletion timescale (left) and { molecular-gas-to-stellar-mass ratio} (right).  In Fig.~\ref{fig:gas_fraction} we show the evolution of the {molecular-gas-to-stellar-mass ratio}. 

{In the left and right panels of Fig.~\ref{fig:gas_properties} the x-axis values $\tau_{\rm dep}$ and $M({\rm H}_2)/M_\star$ have been rescaled by $(1+z)^{\rm B_t}$ and $\eta(z)$, respectively, to remove the redshift dependence, as estimated by \citet{Tacconi2018}, where B$_{\rm t}=-0.62$ and $\log\,\eta(z) = -3.62\times[\log(1+z)-0.66]^2$. }

We have corrected, where needed, the molecular mass estimates from the literature to take the different conversion factors $\alpha_{\rm CO}$ into account. We have assumed $\alpha_{\rm CO}=4.36~M_\odot({\rm K~km}~{\rm s}^{-1}~{\rm pc}^2)^{-1}$, in agreement with the value adopted in this work, while for each galaxy we have assumed the excitation level adopted in the corresponding work.
Similarly to \citet{Castignani2018} we stress that by assuming the same $\alpha_{\rm CO}$ we aim at having comparable molecular gas mass estimates for the galaxies considered. According to our SFR$<6$~SFR$_{\rm MS}$ criterion the majority of the sources lie around the MS, which justifies the choice for $\alpha_{\rm CO}$.

In the figures we have color coded cluster galaxies at $z\lesssim2$ differently from those at $z\gtrsim2$. We also highlight our five targets as well as, to the best of our knowledge, the most distant (proto-)BCGs detected in CO. They are i) the BCG at $z=1.7$ observed in CO(2$\rightarrow$1) by \citet{Webb2017}, ii) { the $z=1.99$ BCG candidate of the cluster Cl~J1449+0856 observed in both CO(4$\rightarrow$3) and CO(3$\rightarrow$2) by \citet{Coogan2018}}, iii) the proto-BCG candidate at $z=2.2$, namely, MRC~1138-262, detected in CO(1$\rightarrow$0) by \citet{Emonts2013,Emonts2016} and, { iv) the proto-BCG candidate Candels-5001 at $z=3.47$ observed in CO(4$\rightarrow$3) by \citet{Ginolfi2017}.}

In addition to Candels-5001 the other two CO detections reported in the figures at redshifts higher than that of COSMOS-FRI~70 and selected to have SFR$<6$~SFR$_{\rm MS}$ are the galaxy GN20.2b at $z=4.0563$ \citep{Hodge2013,Tan2014} and HDF850.1 at $z=5.2$ \citep{Walter2012}.

{We stress here that the BCG candidate of the cluster Cl~J1449+0856 at $z=2$ was first described by \citet{Gobat2011} and has recently been observed with ALMA \citep{Coogan2018,Strazzullo2018}. No stellar mass estimate is reported for this source in these studies. 
However, an absolute K$_s$-band AB magnitude $\sim20$ is reported by \citet{Gobat2011} for the likely interacting triplet of galaxies associated with the proto-BCG. Following \citet{Gobat2011} we assumed a stellar-mass-to-light ratio of 
$M_\star/L_{\rm K_s}=3.8~M_\odot/L_\odot$, where $L_{\rm K_s}$ is the rest frame ${\rm K_s}$-band luminosity in units of the solar luminosity ($L_\odot$). This procedure yielded an estimate of the BCG stellar mass $M_\star=(5.0\pm2.7)\times10^{11}~M_\odot$ which we have used in this work.}

\subsection{Megaparsec-scale overdensities}\label{sec:Mpcscale_overdensities}
{ We searched for megaparsec-scale overdensities around the radio sources of our sample using the $\mathit{w}$PPM cluster-finder procedure and photometric redshifts of galaxies, as described in Sect.~\ref{sec:wPPM}. 

In Figs.~\ref{fig:PPM_plots_DES} and \ref{fig:PPM_plotsCOSMOS} (right) we show the density maps corresponding to the overdensities associated with the radio sources. In the same figures we also show the peak and size ($\mathcal{R}_\mathit{w}$) of the detection as found by the wavelet transform.
We also report in the same figures the PPM plots for the fields of the radio sources. For each pair ($z_{\rm centroid};\Delta z$) we have plotted the detection significance, where different colors correspond to different detection significances. Points with associated significance $<2\sigma$ are not plotted. In each PPM plot the vertical solid line shows the spectroscopic redshift of the radio galaxy. 

The $\mathit{w}$PPM results are summarized in Table~\ref{tab:cluster_properties}. Megaparsec-scale overdensities with estimated core size in the range $\sim$(0.5-1)~Mpc are found for all radio sources in our sample at redshifts consistent with the spectroscopic redshifts of the radio galaxies. }

\begin{table*}[]
\begin{adjustwidth}{-0.5cm}{}
\begin{tabular}{c|cccccccccc}
\hline\hline
 Galaxy ID &  $z_{spec}$ & $({\rm R.A.})_{ov}$ & $({\rm Dec.})_{ov}$ & $\theta_{ov}$ & $z_{ov}$ & significance & $\overline{\Delta z}$ & $N_{\rm selected}$  & $\mathcal{R}_{\rm PPM}$ & $\mathcal{R}_{\mathit{w}}$    \\
 (1) & (2) & (3) & (4) & (5) & (6) & (7) & (8) & (9) & (10) & (11) \\ 
 \hline
{\small DES-RG~399} &    {\small 0.388439} & 40.58809994  & -0.5746418  & 92.1$''$ & 0.34$\pm$0.06   & 2.6$\sigma$  & 0.2 & 6~(1.3) & {100$''$}  &  {175$''$} \\
     &  &  &  & 486~kpc &    &   &  &  & 528~kpc &  923~kpc \\
\hline
 {\small DES-RG~708} &   {\small 0.60573}  &  41.34062917  & -0.54419315 & 9.3$''$  & 0.61$\pm$0.05  & 6.0$\sigma$  & 0.2 & 17~(2.3)   & 150$''$  & 125$''$ \\
      &  &  &  & 62~kpc &    &   &  &  & 1007~kpc &  839~kpc \\
 \hline 
 {\small COSMOS-FRI~16} &  {\small 0.9687}  &  150.54213359  & 2.27617522   & 35.5$''$ & 0.93$\pm$0.06     & 3.1$\sigma$ & 0.2 & 121~(93.0)  & 100$''$   & 107$''$ \\ 
       &  &  &  & 282~kpc &    &   &  &  & 795~kpc &  850~kpc \\
       \hline
                        &                   &  150.50776446  & 2.25025075   & 124.1$''$  & 1.06$\pm$0.06 &  5.5$\sigma$  & 0.2 & 247~(198.9)  & 145$''$  & 155$''$  \\ 
                               &  &  &  & 986~kpc &    &   &  &  & 1152~kpc &  1232~kpc \\
 \hline 
 {\small COSMOS-FRI~31} &  {\small 0.9123}  &   149.63013031 & 1.9426736   & 102.6$''$  & 0.96$\pm$0.05 &  2.3$\sigma$ & 0.2 & 34~(24.5) &    112$''$ & 106$''$ \\ 
                                &  &  &  & 802~kpc &    &   &  &  & 876~kpc &  829~kpc \\
 \hline
 {\small COSMOS-FRI~70} &  {\small 2.625}  & 150.64167779 & 2.29597309 & 82$''$ & 2.65$\pm$0.04 & 2.8$\sigma$ & 0.15 & 7~(1.6)  & 87$''$  & 88$''$  \\
 &  &  &  & 655~kpc &    &   &  &  & 694~kpc &  702~kpc \\

  \hline
\end{tabular}
\caption{Properties of the megaparsec-scale overdensities around the radio sources:  (1-2) galaxy ID and spectroscopic redshift; (3-4) J2000 projected space coordinates of the overdensity peak as found by the wavelet transform; (5) projected separation between the coordinates in columns (3-4) and those of the radio source; overdensity  redshift (6) and  significance (7) as found by the PPM; photometric redshift bin (8) and number of sources (9) selected by the PPM to detect the overdensity, while between parentheses there is corresponding average number of sources in the survey within the same redshift bin and overdensity area; (10-11) radius of the overdensity as found by the PPM and by the wavelet transform. Columns (5,10,11) are reported in units of both arcseconds and kiloparsecs (physical).}
\label{tab:cluster_properties}
\end{adjustwidth}
\end{table*}

\subsubsection{DES-RG~399 and 708}
We searched for overdensities around DES-RG~399 and 708 using photometric redshifts from the DR14 of SDSS, which has an AB magnitude limit of $\textsf{i}\sim21.3$. As a consistency check we verified that the cluster candidates around the two DES-RGs are both detected also when the deeper year~1 DES SN deep-field number~3 photometric redshift catalog (DES collaboration; C.~Benoist, private comm.) is used. DES SN deep field photometry is in fact complete down to AB magnitudes $\textsf{i}\sim24.5$ and allowed us to detect both overdensities around DES-RG~399 and 708 at $z_{ov}=0.31\pm0.05$ and $z_{ov}=0.60\pm0.05$, with higher significance ($3.8\sigma$ and $8.3\sigma)$ than that obtained using SDSS photometric redshifts, respectively. 

%
%
%
%

We  also searched for overdensities associated with DES-RG~399 and 708 within both RedMapper\footnote{http://risa.stanford.edu/redmapper/} \citep[v6.3,][]{Rykoff2014} and \citet{Wen2012} cluster catalogs. Both catalogs were built using the SDSS photometric dataset and include cluster candidates up to $z\simeq0.8$.  Our search did not produce positive results, suggesting that the cluster candidates around DES-RG~399 and 708 might be associated with (rich) groups of more than massive $\gtrsim10^{14}~M_\odot$ clusters. In particular at the redshift of DES-RG~399 ($z=0.39$) the completeness of the \citet{Wen2012} cluster catalog is estimated to be about $\simeq80\%$, while it drops significantly down to $\simeq30\%$ at the redshift of DES-RG~708 ($z=0.61$), for $M_{200}>6\times10^{13}$ cluster masses.\footnote{Here $M_{200}$ is the mass enclosed by a radius encompassing the matter density 200 times the critical one.}

\subsubsection{COSMOS-FRI~16, 31, and 70}
We searched for overdensities associated with COSMOS-FRI~16, 31, and 70 using the $\mathit{w}$PPM and photometric redshifts of galaxies from both \citet{Ilbert2009} and \citet{Laigle2016} photometric redshift catalogs. Overdensities associated with the COSMOS-FRIs are found for all three radio galaxies when the \citet{Ilbert2009} photometric redshift catalog is used. Cluster candidates associated with COSMOS-FRI~16 and 31 are also found when the \citet{Laigle2016} photometric redshift catalog is instead used. However the proto-cluster candidate around COSMOS-FRI~70 is found only when the \citet{Ilbert2009} photometric redshift catalog is adopted. This discrepancy might be due to different photometric selection associated with the \citet{Ilbert2009} and \citet{Laigle2016} catalogs.  In Fig.~\ref{fig:PPM_plotsCOSMOS} and Table~\ref{tab:cluster_properties} we report the results of our analysis for COSMOS-FRI~16 and 31, where the \citet{Laigle2016} catalog has been used, while for COSMOS-FRI~70 the results refer to the \citet{Ilbert2009} catalog.

\subsection{Properties of the galaxies in the overdensities}\label{sec:results_cluster_galaxies}
In this section we aim to characterize the global properties of the galaxy population associated with the megaparsec-scale overdensities found around the radio galaxies in our sample. As outlined in the following a number of limitations prevent us from assigning (proto-)cluster membership using sophisticated probabilistic methods based on photometric redshifts of galaxies \citep[e.g.,][]{Castignani_Benoist2016,George2011}. Cluster membership algorithms are usually applied to large samples of clusters/groups, that is, cluster membership derived with such methods has a statistical meaning. Limiting ourselves to the megaparsec-scale overdensities associated with the radio source in our sample, it is worth also noting that i) the (proto-)cluster structure and topology are complex and that ii) the estimated richness ($N_{\rm selected}$) of the overdensities is relatively low. These two aspects limit us from applying accurate cluster membership assignments that take into account the cluster-centric distance of each galaxy \citep[e.g.,][]{Rozo2009,Rozo2015,Castignani_Benoist2016}. Furthermore defining the proto-cluster galaxies that will end up in a virialized cluster by $z\sim0$ is not a trivial task, as shown by \citet{Contini2016}, based on semi-analytic simulations. More precisely they showed that the fraction of galaxies in the proto-cluster field that are not progenitors of $z\sim0$ cluster galaxies is not negligible, and about $\sim(20-30)\%$ for galaxies with $M_\star\simeq10^9M_\odot$.

For these reasons we preferred to adopt a heuristic but still effective approach to select fiducial (proto-)cluster members, as described in the following. We verified {\it a posteriori} that this approach still gives reasonable and consistent results. Visual inspection of the density maps in Figs~\ref{fig:PPM_plots_DES} and \ref{fig:PPM_plotsCOSMOS} (right) suggests that the overdensity peaks as found by the PPM and its wavelet-based upgrade are delimited within regions of similar size, that is, $\mathcal{R}_{\rm PPM}\simeq\mathcal{R}_{\mathit{w}}$ (see also Table~\ref{tab:cluster_properties}). Furthermore, such regions are reasonably well contained within a circle centered at the radio galaxy coordinates and with a radius of 1~Mpc, which typically defines the cluster core size. We have therefore selected all galaxies within a projected distance of 1~Mpc from each radio source in our sample. The 1-Mpc radius was evaluated at the spectroscopic redshift $z_{\rm RG}$ of the radio source. Among such sources we then selected as fiducial (proto-)cluster members only those that have photometric redshifts within $z_{\rm RG}-\sigma_0(1+z_{\rm RG})$ and $z_{\rm RG}+\sigma_0(1+z_{\rm RG})$, with $\sigma_0=0.03$,  which is consistent with the value adopted for the PPM procedure (see Sect.~\ref{sec:PPM_proc}).

\subsubsection{Color-magnitude and color-color plots}
We made color-magnitude (CM) plots using the galaxies in the field of the radio galaxies, which were selected within a projected radius of 1~Mpc as described in the previous section. In particular we produced \textsf{g-i} versus \textsf{i} CM plots for sources in the field of DES-RG~399 and 708, using SDSS photometry. We also produced \textsf{r-K$_s$} versus \textsf{K$_s$} CM plots for sources in the field of COSMOS-FRI~16, 31, and 70 using available COSMOS \textsf{r}- and \textsf{K$_s$}-band  photometry from the Suprime-CAM of Subaru \citep{Taniguchi2007} and UltraVISTA \citep{McCracken2012}, respectively,  derived from fixed $3''$ aperture, as reported in \citet{Laigle2016}. We note that the \textsf{g}- and \textsf{i}-band filters of SDSS have effective wavelengths equal to 4,770 and 7,625~\AA. They were chosen because they optimally catch the 4,000~\AA~ break of DES-RG~399 and 708, which are redshifted at 5554 and 6423~\AA, respectively. Similarly, the \textsf{r}- and \textsf{K$_s$}-band filters of Subaru Suprime-CAM and UltraVISTA have effective wavelengths equal to 6,289 and 21,540~\AA. The redshifted 4,000~\AA~ breaks of COSMOS-FRI~16, 31, and 70 optimally fall between these two wavelengths, being located at 7,875, 7,649,  and 14,500~\AA, respectively. 

Concerning COSMOS-FRI~16, 31, and 70 we used their Subaru \textsf{r}- and CFHTLS \textsf{K$_s$}-band photometry from B13, consistently with what has been done for the SED modeling in Sect.~\ref{sec:SEDmodeling}. However we verified that using instead the photometric data from \citet{Laigle2016} does not change our final results.

The cluster candidates around DES-RG~399 and 708 have a small number of selected galaxies (see also Table~\ref{tab:cluster_properties}). Therefore, we prefer not to show any CM plot for these overdensities.
However by using the year~1 DES SN deep field number~3 photometric redshift catalog (DES collaboration; C.~Benoist, private comm.) we verified that with the SDSS dataset we are indeed targeting  only the bright end of the CM plot, that is, of the cluster galaxy luminosity function.

\begin{figure*}[h!] \centering
\subfloat{\includegraphics[width=0.46\textwidth]{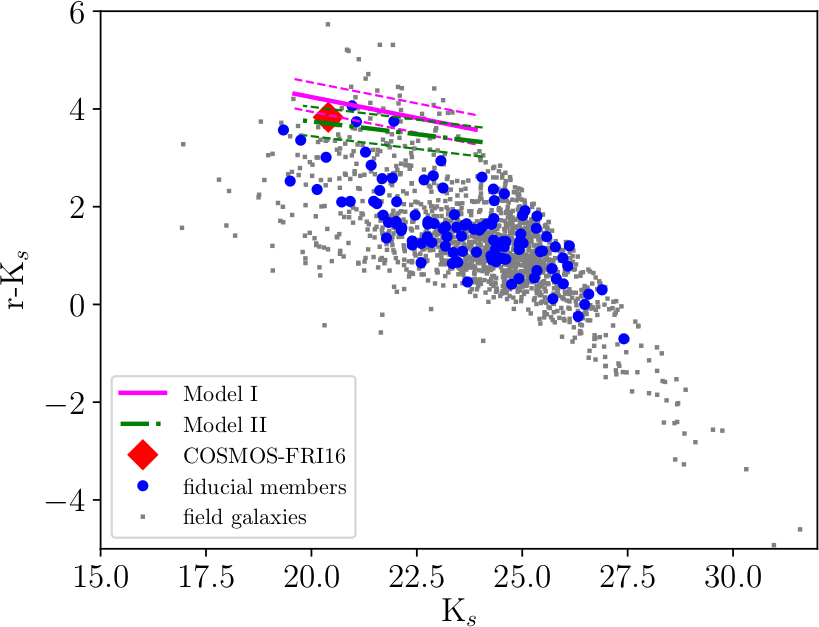}}
\subfloat{\hspace{0.3cm}\includegraphics[width=0.46\textwidth]{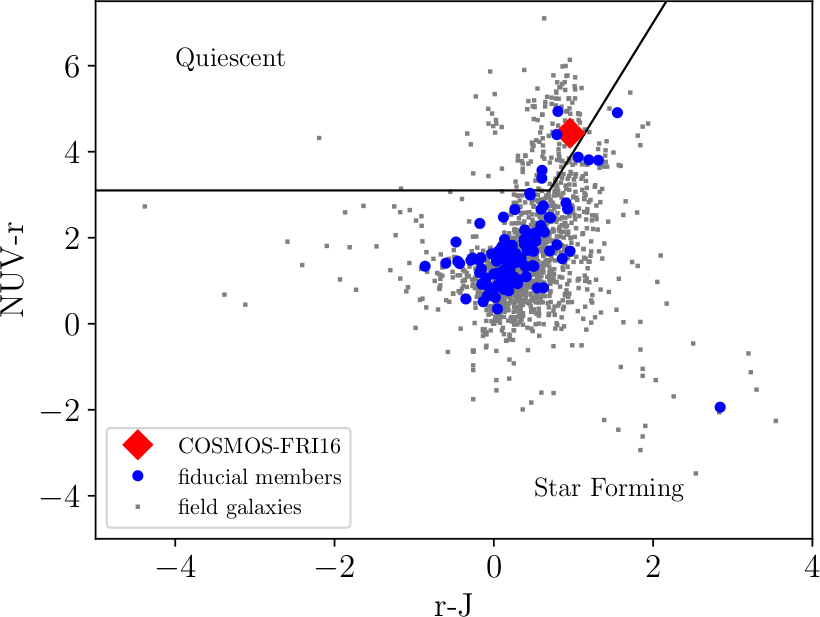}}\qquad
\subfloat{\includegraphics[width=0.46\textwidth]{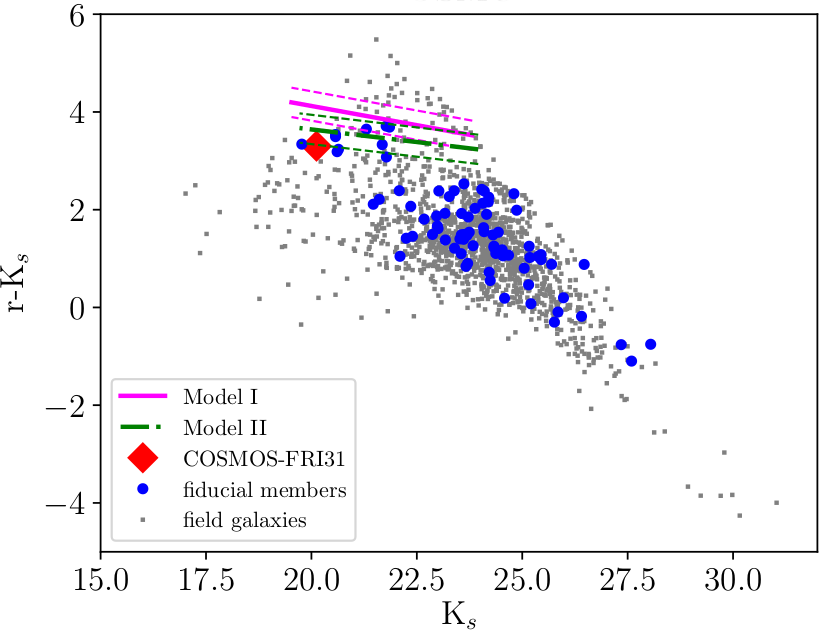}}
\subfloat{\hspace{0.3cm}\includegraphics[width=0.46\textwidth]{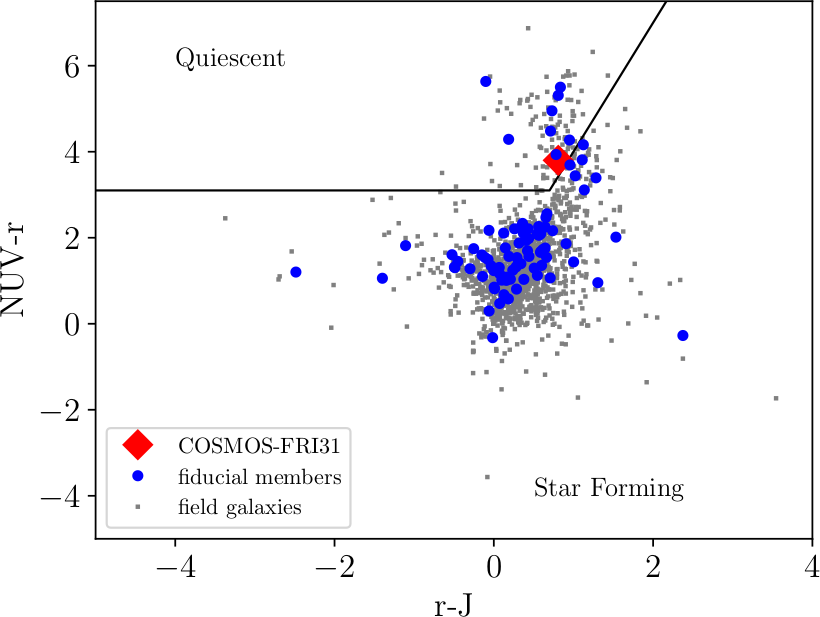}}\qquad
\subfloat{\includegraphics[width=0.46\textwidth]{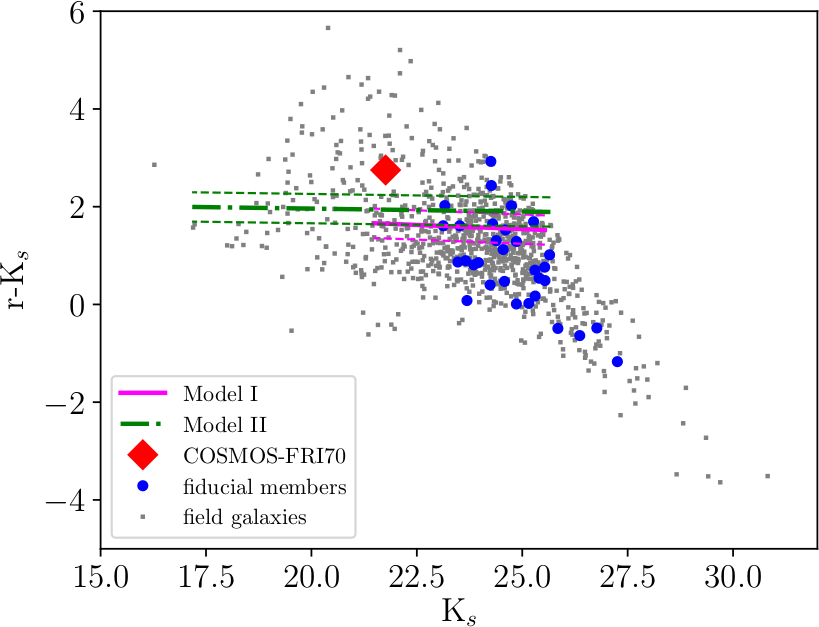}}
\subfloat{\hspace{0.3cm}\includegraphics[width=0.46\textwidth]{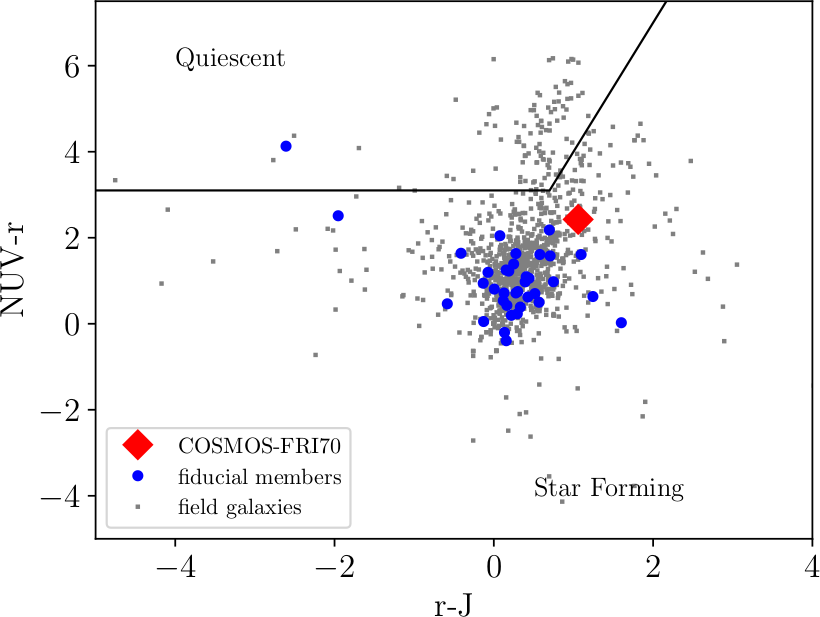}}\qquad
\caption{Color-magnitude  (left) and color-color (right) plots for sources within a projected distance of 1~Mpc  from the COSMOS-FRI~16 (top), 31 (center), and 70 (bottom). Fiducial cluster members and COSMOS-FRI galaxies are distinguished from the remaining sources in the field. Left: Red-sequence models for galaxies between $10^{10}$ and $5\times10^{11}~M_\odot$ are plotted. Red-sequence Models~I and II  correspond to a formation redshift $z_f=6.5$ and 20, respectively. The horizontal dashed lines show a fiducial $\pm0.3$~dex uncertainty in each model, similarly to \citet{Kotyla2016}. Right: The solid black line distinguishes between fiducial star-forming and quiescent galaxies, respectively. See text for further details.}
\label{fig:CM_CC_plots}
\end{figure*}

In Fig.~\ref{fig:CM_CC_plots} (left) we report the CM plots for sources in the field of COSMOS-FRI~16, 31, and 70, where two red-sequence models corresponding to formation redshifts $z_f=20$ and 6.5 are overplotted and estimated using the Galaxy Evolutionary Synthesis Models (GalEv) tool\footnote{http://www.galev.org} \citep{Kotulla2009}. Model parameters are reported in Table~\ref{tab:Galev_par} and are equal to those adopted by \citet{Kotyla2016}, who studied the megaparsec-scale environments of distant radio sources.

\begin{table}
\begin{center}
\begin{tabular}{ccr}
\hline\hline
$M_\star$ & & [Fe/H] \\
($M_\odot$) & &        \\
\hline\hline
&  & \\
& Model~I ($z_f=6.5$) & \\
\hline
$1\times10^{10}$ & & 0.0 \\
$5\times10^{11}$ & & +0.3 \\
\hline\hline
&  & \\
& Model~II ($z_f=20$) & \\
\hline
$1\times10^{10}$ & & -0.3 \\
$5\times10^{11}$ & & 0.0 \\
\hline
\end{tabular}
\end{center}
\caption{GalEv parameters. For each formation redshift $z_f=6.5$, 20, the stellar mass and metallicity used for galaxy evolution model are reported. We also assume the following GalEv input parameters: (1) IMF: Salpeter IMF between (1-100)~$M_\odot$; burst: no burst; type: E (elliptical); and extinction law: none. The reported parameters used in this work for the red-sequence modeling are the same as in \citet{Kotyla2016}.}
\label{tab:Galev_par}
\end{table}

In the right panels of the same figure we also report the rest frame NUV-\textsf{r} versus \textsf{r-J} color-color plots for the same sources in the field of the COSMOS-FRI radio galaxies, as in the left panels. Such diagrams are commonly used in the literature to distinguish fiducial star-forming from quiescent galaxies \citep{Williams2009,Laigle2016}. Here NUV, \textsf{r}, and \textsf{J} denote magnitudes from GALEX, Subaru, and VISTA, respectively, from the \citet{Laigle2016} catalog.

\subsubsection{Morphological classification}

To investigate the morphology of our target COSMOS-FRI radio galaxies, we determined their half-light radius and S\`{e}rsic index from high-resolution HST ACS F814W (I-band) images using Galfit \citep{Peng2002,Peng2010}. The pixel size of these images (see Fig.~\ref{fig:COSMOS_FRI_images}) is 0.03~arcsec, for an effective resolution of about 0.1~arcsec ($\sim$0.8~kpc at the target redshifts). On the other hand we stress here that both SDSS and DES images have a resolution of $\lesssim1$~arcsec which does not allow us to optimally fit the light profile of DES-RG~399 and 708.

The S\`{e}rsic law can be expressed as
\begin{equation}
 \Sigma(r)=\Sigma_e\exp\left[b_n\left(\left(\frac{r}{r_e}\right)^{1/n}-1\right)\right]\;,
\end{equation}
where $\Sigma(r)$ is the surface brightness at a projected distance $r$ from the galaxy center, $\Sigma_e$ is the surface brightness at the half-light radius $r_e$, $n$ is the S\`{e}rsic index, and $b_n$ is a function of the S\`{e}rsic index.
The S\`{e}rsic law with $n=4$ fairly approximates the profile of classical elliptical galaxies, for which it holds $b_4=7.669$ \citep[see e.g.,][and references therein]{Graham_Driver2005}. 

The fits were performed using a generic Tiny Tim \citep{Krist1995,Krist2011} point spread function for the HST ACS F814W filter, assuming a G2V stellar spectrum.
Our Galfit analysis yielded $n=3.7\pm0.2$, $4.4\pm0.1$, and $4.3\pm0.6$, as well as $r_e=(4.9\pm0.8)$, (4.7$\pm$0.8), and (2.9$\pm$0.8)~kpc for COSMOS-FRI~16, 31, and 70, respectively. The estimated S\`{e}rsic indexes are consistent with those of elliptical galaxies ($n=4$). The inferred effective radii $r_e$ are fairly consistent with, albeit smaller than, those predicted by \citet{vanderWel2014} for MS galaxies of similar mass and redshift of COSMOS-FRI~16, 31, and 70, which are equal to 6.6, 6.6, and 4.8~kpc, respectively.

\section{Discussion}\label{sec:discussion}
{ The main goals of this work are to study the molecular gas associated with distant radio galaxies and to characterize their megaparsec-scale environments.} To this aim we targeted CO with the IRAM-30m telescope in five radio galaxies at $z\sim0.4-2.6$, which approximately corresponds to $\sim7$~Gyr of cosmic time. With more statistics, this will contribute to probing the cosmic history of molecular gas reservoirs in early-type (proto-)cluster core galaxies.

\subsection{Molecular gas properties}
\subsubsection{Distant (proto-)BCG candidates with CO observations}
 We observed the five radio galaxies in our sample with the IRAM-30m telescope, targeting several CO(J$\rightarrow$J-1) lines, one per radio source and {set upper limits to the total molecular gas mass for all five sources. However, a hint at $2.2\sigma$ of a possible detection of CO(7$\rightarrow$6) is found for COSMOS-FRI~70, at $z=2.6$. If confirmed with higher-S/N data it would become one of the most distant proto-BCG candidates detected in CO so far.} In general, as further discussed in Sect~\ref{sec:discussion_overdensities},  the present study increases the statistics of CO observations in distant (proto-)BCG candidates.


In fact, as further mentioned in Sect.~\ref{sec:molecular_gas_diagrams}, 
to the best of our knowledge there are only a few distant (proto-)BCGs observed in CO
with associated stellar-mass estimates. They are i) the cluster galaxy at $z=1.7$ observed by \citet{Webb2017}, ii) { the BCG candidate at $z=1.99$ of the cluster Cl~J1449+0856 \citep{Coogan2018}}, iii) MRC~1138-262 at $z=2.2$ from \citet{Emonts2013,Emonts2016}, and iv) { Candels-5001 at $z=3.472$ \citep{Ginolfi2017}.} 

{ We point out here that our five distant radio galaxies were selected from among the ones with the strongest evidence of ongoing star formation within a total of $\sim$30~deg$^2$ comprising the COSMOS survey and the DES SN deep fields 2 and 3. Therefore, although distant star forming BCGs have been reported in previous studies \citep{Webb2015,McDonald2016,Bonaventura2017}, our results suggest that large molecular gas reservoirs associated with them are rare.}

\subsubsection{Main sequence}
{The upper limits to the molecular gas content  
$M({\rm H}_2)/M_\star<$0.11, 0.09, 1.8, 1.5, and 0.29 for the five sources in our sample at $z=0.4$, 0.6, 0.91, 0.97, and 2.6, respectively, are consistent with the empirical values} by \citet{Tacconi2018} for MS field galaxies.
Similarly, the five radio galaxies have estimated depletion time {upper limits in the range $\tau_{\rm dep}\lesssim(0.2-7)$~Gyr} that are consistent with the values of MS galaxies, given the large uncertainties associated with both molecular gas mass and SFR estimates (see also Table~\ref{tab:radio_galaxies_properties_mol_gas} and Fig.~\ref{fig:gas_properties}). 


We note that we assumed a Galactic $\alpha_{\rm CO}$ conversion factor for all our target sources, typical of MS galaxies. Assuming a lower $\alpha_{\rm CO}$, typical of starburst galaxies \citep[e.g.,][for a review]{Bolatto2013}, would lead to a lower molecular gas mass and a shorter depletion timescale. 

The fact that our target sources are consistent with being on the MS suggest that they are not yet merging with their nearby galaxies. 
Concerning  COSMOS-FRI~70, the spectrum reported in Fig.~\ref{fig:IRAM30m_results} implies an estimated FWHM$=(225\pm123)$~km for the CO(7$\rightarrow$6) line. This FWHM supports the scenario where one single source is responsible for the observed {tentative} emission, despite the presence of nearby companions within the IRAM-30m beam (Fig.~\ref{fig:IRAM30m_results}, right). 

However, the  possibility of an interaction of the radio galaxies with their companions cannot be excluded, which is in agreement with previous studies that found that radio-loud AGNs are indeed in systems that are close to merging \citep{Chiaberge2015}. The interaction may lead to a starburst phase, as found for example in distant BCGs \citep[$z\simeq1.7$,][]{Webb2017} as well as local major merging gas-rich pairs such as the Antennae Galaxies \citep{Gao2001,Ueda2012,Whitmore2014}. We also point out that we cannot exclude the presence of projection effects.  
We stress that deeper observations are needed to confirm the CO(7$\rightarrow$6) emission and the line properties.

\subsection{Megaparsec-scale overdensities around low-luminosity radio galaxies}\label{sec:discussion_overdensities}
In this section we discuss the properties of the megaparsec-scale environments associated with the radio galaxies in our sample. We show that our targets are indeed consistent with being among the most distant BCGs with constraints on their molecular gas mass. 

{ Overdensities associated with all five radio galaxies in our sample were found by applying the $\mathit{w}$PPM, which uses the photometric redshift of galaxies.} 

\subsubsection{Miscentering of the radio galaxies and morphology of the overdensities }
Our wavelet-based analysis suggests that the overdensities have a complex morphology. DES-RG~708 at $z=0.61$ is the only radio source which is found to be coincident in the projected space with its associated overdensity peak, namely a negligible 62-kpc miscentering is found. For the other four radio galaxies, a miscentering in the range $\sim(0.3-1)$~Mpc is found, which suggests that the cluster candidates might not be fully relaxed or may still be forming. Interestingly, \citet{Ledlow_Owen1995} found that $\sim90\%$ of the radio galaxies in local ($z<0.09$) Abell clusters are within $\sim200$~kpc from the cluster center. Similarly,  \citet{Smolcic2011} found that LLRGs with $L_{\rm 1.4~GHz}\simeq10^{30.6-32}$~erg~s$^{-1}$~Hz$^{-1}$ are preferentially located within $0.2\times r_{200}$, that is, within $\sim60$~kpc from the center of X-ray selected groups up to $z\sim1.3$.

{ In \citet{Castignani2014} we searched for overdensities around the parent sample of  approximately 30 COSMOS-FRI radio galaxies from \citet{Chiaberge2009}. We also used photometric redshifts from \citet{Ilbert2009}. We found that some radio sources show evidence for a miscentering $\lesssim120$~kpc, while others show offsets of up to $\sim500$~kpc. The offsets were estimated comparing the projected coordinates of the COSMOS-FRIs with those of the fiducial centers of clusters and groups found in previous studies on the basis of X-ray emission and photometric/spectroscopic redshifts of galaxies \citep{Finoguenov2007,George2011,Knobel2009,Knobel2012,Diener2013}. }

In particular, the overdensities associated with COSMOS-FRI~16 and 31 are found in the \citet{Knobel2012} and \citet{Knobel2009} group catalogs, respectively, at an angular separation of 63 and 15~arcsec.  Such offsets are not in full agreement with those found by the $\mathit{w}$PPM using photometric redshifts (see Table~\ref{tab:cluster_properties}). We suggest that the discrepancy might be due to the different datasets used to select the cluster candidates.{  \citet{Knobel2009,Knobel2012} used spectroscopic redshifts from the zCOSMOS-bright survey \citep{Lilly2007} to search for overdensities.} Such a discrepancy implies also that precise assessment of  the center of the overdensity  is in general difficult, especially in the case of low-richness systems, as in the case of our target galaxies (see Sect.~\ref{sec:discussion_lowrichness}).

{ We stress here that in the present work i) we have adapted the PPM to the specific case where the spectroscopic redshifts of the target radio galaxies are available (see Sect.~\ref{sec:PPM_proc}) and ii) we have improved the PPM cluster finder by using a wavelet-based approach as described in Sect.~\ref{sec:wPPM_upgrade}. Therefore the results presented in this work supersede those of \citet{Castignani2014}, concerning COSMOS-FRI~16, 31, and 70. In the following we perform a comparison between such results. }

\subsubsection{Comments on the overdensity detections for COSMOS-FRI~16, 31, and 70}
Limiting ourselves to the three COSMOS-FRI sources of this work, in \citet{Castignani2014} we only found an overdensity, associated with COSMOS-FRI~16, at $z_{ov}=1.12\pm0.06$. The reason why no overdensity was found to be associated with the other two radio sources is that in \citet{Castignani2014} we applied a smoothing procedure which allowed us to remove noisy features but also filtered-out low-S/N overdensities. This can be seen from visually inspecting the PPM plots of COSMOS-FRI~16, 31, and 70 shown in Appendix~A of \citet{Castignani2014_thesis}. Low-S/N features, that is, those at $\gtrsim2\sigma$, are visible in the PPM plots in Figs.~\ref{fig:PPM_plots_DES} and \ref{fig:PPM_plotsCOSMOS}, but they were conservatively not identified as cluster candidates by the PPM peak finding procedure in \citet{Castignani2014} and \citet{Castignani2014_thesis}. The PPM was in fact originally conceived and applied to deal with photometric redshifts of the target radio galaxies, which increase the chance of having projection effects. 

In this work we have optimized the PPM to the case where the spectroscopic redshifts of the radio galaxies are available. This was done mainly by i) applying no filtering procedure and ii) searching for an optimal redshift bin $\overline{\Delta z}$ at which the overdensity is detected in the PPM plot (see Sect.~\ref{sec:PPM_proc}). Such improvements ultimately allowed us to effectively find overdensities around all three COSMOS-FRI sources in our sample.
We also stress that COSMOS-FRI~16 is the only radio source in our sample for which two overdensities are detected in its field at a redshift consistent with that of the radio galaxy. For the sake of completeness both overdensities are shown in Fig.~\ref{fig:PPM_plotsCOSMOS} and  their estimated properties are reported in Table~\ref{tab:cluster_properties}.

\subsubsection{Low-richness structures}\label{sec:discussion_lowrichness}
As mentioned above, in \citet{Castignani2014} we also searched for overdensities associated with COSMOS-FRI sources using existing catalogs of clusters and groups. 
Limiting ourselves to COSMOS-FRI~16 and 31  at $z<1$,  the associated cluster candidates were found within the \citet{Knobel2012} and \citet{Knobel2009} group catalogs, respectively, with estimated masses of $1.9\times10^{13}~M_\odot$ and $8.9\times10^{12}~M_\odot$. 

The overdensity found with the $\mathit{w}$PPM around COSMOS-FRI~70 at $z=2.625$ likely belongs to a (proto-)cluster. To the best of our knowledge this proto-cluster candidate has never been reported in previous work. If confirmed spectroscopically, it will be one of the most distant proto-clusters discovered around radio galaxies \citep[see also e.g.,][]{Noirot2016,Noirot2018}.  \citet{Castignani2014} searched for overdensities associated with $z>1$ COSMOS-FRI radio galaxies within the proto-cluster catalog of \citet{Diener2013}, which was constructed using spectroscopic redshift information from the zCOSMOS-deep survey \citep[][Lilly et al., in prep.]{Lilly2007}. We found only one potential match, that is, COSMOS-FRI~03. 
In \citet{Castignani2014} we also looked for cluster candidates using the \citet{Papovich2008} method, which is based on overdensities of sources detected at 3.6$\mu$m and 4.5$\mu$m and is suited to search for (proto-)clusters at  $z>1.3$ . Among the three COSMOS-FRI sources considered in this work, COSMOS-FRI~70 is the only one at $z>1.3$. We did not find any overdensity associated with such a radio source using the \citet{Papovich2008} method, which suggests that the overdensity around COSMOS-FRI~70 found by the $\mathit{w}$PPM might be consistent with those of rich  groups. 

Therefore, based on the results presented here, the megaparsec-scale overdensities around the LLRGs in our sample are likely associated with rich groups rather than massive $\gtrsim10^{14}~M_\odot$ clusters. Such a statement is
also supported by the richness estimates provided by the $\mathit{w}$PPM. About $N_{\rm selected}\sim$(6-34) sources are in fact selected by the $\mathit{w}$PPM to detect the overdensities, with the only exception represented by the overdensity around COSMOS-FRI~16, for which $\gtrsim$100 sources are selected. 


We stress that spectroscopic follow-ups are ultimately needed to confirm the (proto-)cluster candidates around the radio sources in our sample and fully characterize their cluster population.
Nevertheless, in the following section we discuss the properties of the galaxy population of our (proto-)cluster candidates using photometric redshift information.

\subsubsection{Star forming (proto-)BCG candidates}\label{sec:BCG_candidates}
We derived color-color and color-magnitude (CM) plots for galaxies within a projected distance of 1~Mpc from the radio galaxies (see Sect.~\ref{sec:results_cluster_galaxies} and Fig.~\ref{fig:CM_CC_plots}). The colors and magnitudes of the three COSMOS-FRIs considered in this work are fairly consistent with model predictions for a red sequence of elliptical galaxies, independently of the formation redshift ($z_f=6.5$, 20) associated with red-sequence models. COSMOS-FRI~70 has an \textsf{r-K$_s$} color that is formally redder than that predicted for the red sequence. However, the discrepancy might be still within the associated (large) model uncertainties, considering also the relatively high redshift $z=2.625$ of COSMOS-FRI~70. In fact red-sequence modeling of distant (proto-)cluster galaxies might lead to results of which  interpretation is difficult, even with the use of spectroscopic redshifts. { We refer for example to \citet{Kotyla2016} and \citet{Noirot2016} who derived CM plots for galaxies around distant radio-loud AGNs in dense megaparsec-scale environments.}

The three COSMOS-FRIs of this study  are among the brightest galaxies in the corresponding (proto-)clusters. In particular COSMOS-FRI~70 is $\sim0.3$~mag brighter in the \textsf{K$_s$} band than the second-brightest fiducial proto-cluster member, within the 1~Mpc radius. These findings, associated with the high stellar masses, $\log(M/M_\odot)=10.9-11.5$, { and the estimated S\`{e}rsic indices, $n\simeq3.7-4.4$, } suggest that our target radio galaxies might be indeed associated with the (proto-)BCGs of the cluster candidates, or at least giant ellipticals in the cores of the clusters. Such a statement is also supported by the presence of nearby companions of our target radio galaxies,  clearly visible in particular for DES-RG~399, DES-RG~708, and COSMOS-FRI~70 (Figs.~\ref{fig:DES_RG_images}, \ref{fig:IRAM30m_results}). 

The rest frame NUV-\textsf{r} versus \textsf{r-J} color-color plots of the fiducial cluster members suggest that the great majority of the selected members around the COSMOS-FRIs show on-going star formation. This statement applies also to our target radio galaxies. COSMOS-FRI~16 and 31 are formally at the edge of the dividing line separating quiescent from star-forming galaxies, while COSMOS-FRI~70 is safely classified as a star-forming galaxy, according to the color-color plots.
These results are consistent with those found by the SED modeling, which suggests that the three COSMOS-FRIs are indeed star forming, on the basis of their FIR and UV emission. The suggested star formation associated with the majority of the fiducial cluster galaxies, including the target radio galaxies, is not in contradiction with the possible presence of the red sequence, as suggested by the CM plots. In fact red but nevertheless star forming distant BCGs have been suggested also in recent studies \citep{Bonaventura2017}.

\section{Summary and conclusions}\label{sec:conclusions}
{ We investigated the role of dense megaparsec-scale environment in processing molecular gas in LLRGs in the cores of galaxy (proto-)clusters. To this aim we selected a sample of five radio galaxies in the range $z=0.4-2.6$ 
within the COSMOS survey and DES SN deep field number~3, as part of a search for distant radio galaxies with evidence of ongoing star formation on the basis of the FIR (WISE or Spitzer MIPS) emission at $\sim24~\mu$m. The radio sources have rest frame 1.4~GHz radio powers that are fairly consistent with those of LLRGs.

We targeted CO in the five LLRGs with IRAM-30m and {set upper limits on the total molecular gas mass for  all five sources} at $z=0.39$, 0.61, 0.91, 0.97, and 2.6.
{We found $M({\rm H}_2)/M_\star<$0.11, 0.09, 1.8, 1.5, and 0.29, respectively.}
{For the most distant radio source, COSMOS-FRI~70 at $z=2.6$, a hint of a tentative CO(7$\rightarrow$6) emission is found at $2.2\sigma$.}

{The upper limits to the molecular gas mass and the depletion time, $\tau_{\rm dep}\lesssim(0.2-7)$~Gyr, for the five LLRGs are in agreement with the values found for MS field galaxies \citep{Tacconi2018} of similar mass and redshift.} 

We assembled and modeled the IR-to-UV SEDs for our targets and estimated their SFR and stellar mass in the range $\simeq(21-245)~M_\odot$/yr and $\log(M_\star/M_\odot)\simeq(10.9-11.5)$, respectively, which are fairly consistent with {empirical values} by \citet{Speagle2014} for MS field galaxies.  

We also searched for megaparsec-scale overdensities associated with the LLRGs using the Poisson Probability Method \citep[PPM,][]{Castignani2014b} and photometric redshifts of galaxies. The PPM has been upgraded in this work with an approach based on the wavelet-transform ($\mathit{w}$PPM). Overdensities are found for all five radio galaxies, at redshifts in agreement with the spectroscopic redshifts of the radio sources. 

Our wavelet-based analysis suggests that the morphology of the megaparsec-scale overdensities is complex, typical of still-forming proto-clusters and not fully relaxed clusters.
Our $\mathit{w}$PPM analysis, and the cross-matching of the LLRGs with existing cluster/group catalogs, also suggest that the detected overdensities are associated with rich ($\lesssim10^{14}M_\odot$) groups. 

Color-color and color-magnitude plots suggest that the LLRGs are star forming and on the high-luminosity tail of the red sequence. Our morphological analysis based on the HST ACS images (F814W filter) of the three COSMOS-FRIs in our sample at $z=0.91$, 0.97, and 2.63 implies S\`{e}rsic indices in the range $n\simeq3.7-4.4$, which are typical of classical ellipticals and strengthen the proposed red-sequence scenario.

Our findings are in overall agreement with a scenario where the five LLRGs in our sample are (proto-)BCG candidates.
The present study thus increases the still limited statistics of early-type galaxies in the cores of distant (proto-)clusters with CO observations. {In particular, COSMOS-FRI 70 at $z=2.6$, becomes one of the most distant BCG candidates with CO observations,} while its associated proto-cluster candidate reported in this study is one of the most distant proto-clusters discovered around radio galaxies.

Higher-resolution and higher-surface-brightness-sensitivity observations will allow us to spatially resolve the radio galaxies in our sample from their nearby companions, to possibly distinguish between different gas-processing mechanisms {(e.g., galaxy harassment, strangulation, ram-pressure, or tidal stripping)}, and to put tighter constraints on their molecular gas properties.
The radio galaxies studied in this work are therefore excellent targets for ALMA as well as next-generation telescopes such as the {\it James Webb Space Telescope.} }

\begin{acknowledgements}
This work is based on observations carried out under project numbers 073-16 and 074-17 with the IRAM 30m telescope. IRAM is supported by INSU/CNRS (France), MPG (Germany) and IGN (Spain). { We thank the anonymous referee for helpful comments that led to a substantial improvement of the paper.} GC thanks Micol Bolzonella for helpful discussion about zCOSMOS-deep sources. { GC and MC thank Colin Norman for the suggestion of using wavelet transforms in the PPM.} 
{ PS acknowledges support from the ANR grant LYRICS (ANR-16-CE31-0011).}
This publication has made use of data products from the Infrared Astronomical Satellite (IRAS), Wide-field Infrared Survey Explorer (WISE), the UKIRT Infrared Deep Sky Survey (UKIDSS), the Sloan Digital Sky Survey (SDSS), the Dark Energy Survey (DES), the COSMOS survey, and the NASA/IPAC Extragalactic Database (NED).
The WISE is a joint project of the University of California, Los Angeles, and the Jet Propulsion Laboratory/California Institute of Technology, funded by the National Aeronautics and Space Administration.
Funding for the Sloan Digital Sky Survey IV has been provided by the Alfred P. Sloan Foundation, the U.S. Department of Energy Office of Science, and the Participating Institutions. SDSS-IV acknowledges support and resources from the Center for High-Performance Computing at the University of Utah. The SDSS web site is www.sdss.org. SDSS-IV is managed by the Astrophysical Research Consortium for the Participating Institutions of the SDSS Collaboration including the Brazilian Participation Group, the Carnegie Institution for Science, Carnegie Mellon University, the Chilean Participation Group, the French Participation Group, Harvard-Smithsonian Center for Astrophysics, Instituto de Astrof\'isica de Canarias, The Johns Hopkins University, Kavli Institute for the Physics and Mathematics of the Universe (IPMU) / University of Tokyo, Lawrence Berkeley National Laboratory, Leibniz Institut f\"ur Astrophysik Potsdam (AIP),  Max-Planck-Institut f\"ur Astronomie (MPIA Heidelberg), Max-Planck-Institut f\"ur Astrophysik (MPA Garching), Max-Planck-Institut f\"ur Extraterrestrische Physik (MPE), 
National Astronomical Observatories of China, New Mexico State University, New York University, University of Notre Dame, 
Observat\'ario Nacional / MCTI, The Ohio State University, Pennsylvania State University, Shanghai Astronomical Observatory, 
United Kingdom Participation Group,Universidad Nacional Aut\'onoma de M\'exico, University of Arizona, University of Colorado Boulder, University of Oxford, University of Portsmouth, University of Utah, University of Virginia, University of Washington, University of Wisconsin, Vanderbilt University, and Yale University.
Funding for the DES Projects has been provided by the U.S. Department of Energy, the U.S. National Science Foundation, the Ministry of Science and Education of Spain, the Science and Technology FacilitiesCouncil of the United Kingdom, the Higher Education Funding Council for England, the National Center for Supercomputing Applications at the University of Illinois at Urbana-Champaign, the Kavli Institute of Cosmological Physics at the University of Chicago, the Center for Cosmology and Astro-Particle Physics at the Ohio State University, the Mitchell Institute for Fundamental Physics and Astronomy at Texas A\&M University, Financiadora de Estudos e Projetos, Funda{\c c}{\~a}o Carlos Chagas Filho de Amparo {\`a} Pesquisa do Estado do Rio de Janeiro, Conselho Nacional de Desenvolvimento Cient{\'i}fico e Tecnol{\'o}gico and the Minist{\'e}rio da Ci{\^e}ncia, Tecnologia e Inova{\c c}{\~a}o, the Deutsche Forschungsgemeinschaft, and the Collaborating Institutions in the Dark Energy Survey.
The Collaborating Institutions are Argonne National Laboratory, the University of California at Santa Cruz, the University of Cambridge, Centro de Investigaciones Energ{\'e}ticas, Medioambientales y Tecnol{\'o}gicas-Madrid, the University of Chicago, University College London, the DES-Brazil Consortium, the University of Edinburgh, the Eidgen{\"o}ssische Technische Hochschule (ETH) Z{\"u}rich,  Fermi National Accelerator Laboratory, the University of Illinois at Urbana-Champaign, the Institut de Ci{\`e}ncies de l'Espai (IEEC/CSIC), the Institut de F{\'i}sica d'Altes Energies, Lawrence Berkeley National Laboratory, the Ludwig-Maximilians Universit{\"a}t M{\"u}nchen and the associated Excellence Cluster Universe, the University of Michigan, the National Optical Astronomy Observatory, the University of Nottingham, The Ohio State University, the OzDES Membership Consortium, the University of Pennsylvania, the University of Portsmouth, SLAC National Accelerator Laboratory, Stanford University, the University of Sussex, and Texas A\&M University.
Based in part on observations at Cerro Tololo Inter-American Observatory, National Optical Astronomy Observatory, which is operated by the Association of Universities for Research in Astronomy (AURA) under a cooperative agreement with the National Science Foundation.
The NED is operated by the Jet Propulsion Laboratory, California Institute of Technology, under contract with the National Aeronautics and Space Administration
\end{acknowledgements}

\end{document}